\documentclass[12pt]{article}%
\usepackage{amssymb,bbm,bm}
\usepackage{amsfonts}
\usepackage{amsmath,calrsfs}
\usepackage{boondox-cal}
\usepackage{tikz}
\usepackage{natbib}
\renewcommand{\cite}{\citet}
\usepackage[nohead]{geometry}
\usepackage[singlespacing]{setspace}
\usepackage[bottom]{footmisc}
\usepackage{indentfirst}
\usepackage{endnotes}
\usepackage{graphicx}%
\usepackage{rotating}
\usepackage{verbatim}
\usepackage{setspace}
\usepackage{multirow}
\usepackage{latexsym}
\usepackage{bigints}

\usepackage{amsthm}
\usepackage{makeidx}
\usepackage{fancyhdr,subcaption}
\usepackage{type1cm}
\usepackage{hyperref}
\usepackage{array,multirow,makecell}

\usepackage[linesnumbered,ruled,vlined]{algorithm2e}
\SetKwInput{KwInput}{Inputs}
\SetKwInput{KwOutput}{Step 2}

\SetKwFunction{Forecasting}{Forecasting}
\usepackage{algpseudocode}

\usepackage{xcolor}
\usepackage{framed}
\colorlet{shadecolor}{yellow}
\definecolor{dgreen}{rgb}{0,0.5,0}
\definecolor{dblue}{rgb}{0,0,0.9}
\definecolor{dred}{rgb}{0.6,0.0,0.1}
\definecolor{dgold}{rgb}{0.5,0.3,0.0}
\definecolor{dvio}{rgb}{0.6,0.3,0.5}
\definecolor{gray}{rgb}{0.5,0.5,0.5}
\definecolor{dbraun}{rgb}{.5,0.2,0}

\newcommand{\beq}{\begin{eqnarray*}}
\newcommand{\eeq}{\end{eqnarray*}}
\newcommand{\EE}{\textbf{E}}
\newcommand{\btheta}{\boldsymbol{\theta}}
\newcommand{\bw}{\textbf{w}}
\newcommand{\by}{\textbf{y}}
\newcommand{\bz}{\textbf{z}}

\newcommand{\bZ}{\textbf{Z}}
\newcommand{\mfm}{\mathfrak{m}}
\newcommand{\st}{\star}
\newtheorem{thm}{Theorem}[section]
\newcommand{\1}{\mathbbm{1}}

\newcommand{\wh}{\widehat}
\newcommand{\wtl}{\widetilde}
\newcommand{\ol}{\overline}
\newcommand{\ul}{\underline}

\newtheorem{defn}{Definition}[section]

\newtheorem{prop}{Proposition}[section]
\newtheorem{assum}{Assumption}[section]

\numberwithin{equation}{section}

\theoremstyle{definition}

\makeatletter
\def\@biblabel#1{\hspace*{-\labelsep}}
\makeatother
\usepackage{xr}
\makeindex

\begin{document}
\newgeometry{lmargin=1.20in,rmargin=1.20in,tmargin=0.7in,bmargin=1.20in}

\title{\huge Panel data models with randomly generated groups}

\author{Jean-Pierre Florens\thanks{
Toulouse School of Economics, Universit\'{e} Toulouse Capitole, Toulouse (France). e-mail: \texttt{jean-pierre.florens@tse-fr.eu}} \and
Anna Simoni\thanks{
CREST, CNRS, \'{E}cole Polytechnique, ENSAE, 5 Avenue Henry Le Chatelier, 91120 Palaiseau (France). Phone: +33(0)170266837. e-mail: \texttt{anna.simoni@polytechnique.edu} (corresponding author)}
}

\maketitle

\begin{abstract}
\begin{singlespace}
We develop a structural framework for modeling and inferring unobserved heterogeneity in dynamic panel-data models. Unlike methods treating clustering as a descriptive device, we model heterogeneity as arising from a latent clustering mechanism, where the number of clusters is unknown and estimated. Building on the mixture of finite mixtures (MFM) approach, our method avoids the clustering inconsistency issues of Dirichlet process mixtures and provides an interpretable representation of the population clustering structure. We extend the Telescoping Sampler of Fruhwirth-Schnatter et al. (2021) to dynamic panels with covariates, yielding an efficient MCMC algorithm that delivers full Bayesian inference and credible sets. We show that asymptotically the posterior distribution of the mixing measure contracts around the truth at parametric rates in Wasserstein distance, ensuring recovery of clustering and structural parameters. Simulations demonstrate strong finite-sample performance. Finally, an application to the income–democracy relationship reveals latent heterogeneity only when controlling for additional covariates.
\end{singlespace}
\end{abstract}

\begin{singlespace}
\end{singlespace}

\section{Introduction}
Understanding individual heterogeneity is essential for analyzing the behavior of economic agents and assessing the impact of economic policies. Economic actors are inherently diverse: no two agents are identical, and their observed and unobserved characteristics shape how they respond to incentives, shocks, and policy interventions. For example, in labor economics, latent traits such as motivation or adaptability may determine how a job seeker benefits from a training program. In macroeconomics, forecasts can be biased if they neglect idiosyncratic features of individual series that cannot be explained by observable covariates.\\
\indent A fundamental distinction must be drawn between observed heterogeneity -- variation explained by observable characteristics such as age, education, or firm size -- and unobserved heterogeneity, which arises from latent attributes. Both types matter for economic modeling, but the unobserved component is particularly challenging because it is not directly measurable. Ignoring such latent heterogeneity can lead to biased estimates, misleading inferences, and flawed policy recommendations.\\
\indent
Panel-data models provide a natural framework for incorporating unobserved heterogeneity by introducing unit-specific time-invariant latent variables. By exploiting repeated observations on the same units over time we can learn about them. These latent features capture persistent differences across individuals, firms, or countries. Importantly, such heterogeneity often has a clustering structure: the population may be partitioned into a finite number of groups, each with distinct characteristics. Detecting and characterizing these clusters is crucial for understanding policy impacts and improving forecasts.\\
\indent This paper develops a new structural approach for modeling and inferring the clustering structure of unobserved heterogeneity in dynamic panel data models. Unlike traditional methods, we do not treat clustering as a purely descriptive device. Instead, we explicitly model the probabilistic mechanism generating the clusters and infer its structure from the data. A key advantage of our approach is that the number of clusters need not be fixed in advance. Instead, it is treated as an unknown parameter, estimated jointly with other structural features of the model. Formally, our framework is based on a mixture of finite mixtures (MFM) model (\cite{RichardsonGreen1997}), where the population distribution of latent features is represented as a finite mixture with an unknown number of components. This approach offers two important advantages. First, it avoids the well-known inconsistency issues of Dirichlet Process Mixture models, in detecting clusters, when the true number of clusters is finite. Second, it provides a flexible yet interpretable representation of heterogeneity, where the population clustering mechanism is characterized by three parameters: the number of groups $K^{\st}$, the latent features (atoms) $\btheta^{\st}$, and their weights $\bw^{\st}$.\\
\indent In addition to latent heterogeneity, the dynamic panel data model that we consider includes the lagged dependent variable and exogenous covariates among the explanatory variables, both of which have homogeneous effects across units. These effects are denoted by $\gamma^{\st}$ and $\beta^{\st}$, respectively. We estimate the parameters $(\gamma^{\st},\beta^{\st},\btheta^{\st},\bw^{\st},K^{\st})$ by combining panel data (with $N$ units and $T$ time periods) with informative priors. The random parameters associated with the true model parameters are denoted by $(\gamma,\beta,\btheta_K,\bw_K,K)$ and are endowed with a prior distribution. To perform inference, we extend the Telescoping Sampling algorithm of \cite{FruwirthSchnatter2021} to accommodate the panel structure, unobserved heterogeneity, exogenous covariates, and lagged dependent variables. This algorithm is computationally efficient, automatically produces credible sets for all parameters, and scales well with both the cross-sectional and time dimensions of the data.\\
\indent Our contributions can be summarized as follows. First, we provide a structural modeling of clustering. We introduce a principled approach to modeling unobserved heterogeneity in panel data as a structural clustering mechanism. Unlike previous work (\textit{e.g.}, \cite{BonhommeManresa2015}), which treats clustering as a descriptive tool without modeling the underlying probabilistic mechanism, we adopt a structural approach and estimate the underlying latent structure. Second, we estimate the number of clusters. We treat the number of clusters as an unknown parameter, avoiding the need for ad hoc choices. We study the role of priors on $K$, showing how the effective number of clusters represented in finite samples (denoted by $K_{+,N}^{\st}$) can differ from the true number of clusters $K^{\st}$, and how this gap vanishes as $N$ grows.\\
\indent Third, we provide theoretical guarantees. We establish identification of the model and demonstrate asymptotic results for the posterior distribution of the mixing measure as the number of units $N$ increases. Specifically, we show that the posterior contracts around the true latent distribution at nearly parametric rates (up to a logarithmic factor), with convergence measured in Wasserstein distance. This ensures recovery of the cluster locations $\btheta^{\st}$, their weights $\bw^{\st}$, the number of clusters $K^{\st}$, and the structural parameters $\gamma^{\st}$, $\beta^{\st}$. Notably, we do not require that $T$ grows to infinity to recover $K^{\st}$ as \textit{e.g.} in \cite{BaiNg2002} and \cite{BonhommeManresa2015}.\\
\indent Fourth, the paper supplies an efficient computation Markov Chain Monte Carlo (MCMC) algorithm. We propose an extension of the Telescoping Sampler of \cite{FruwirthSchnatter2021} tailored to panel data models, which delivers fast and reliable inference. The method provides not only point estimates but also full uncertainty quantification for both clustering and regression parameters. Fifth, we analyse the finite sample performance of our approach and show, through Monte Carlo simulations, that our approach works well in finite samples. When clusters are well-separated, the structure is recovered almost perfectly. In more challenging cases with many clusters or almost overlapping features, larger samples or longer panels help disentangle the heterogeneity. This shows the usefulness of panel data, over cross-section data, to recover the clustering mechanism. In all cases, inference for $\gamma^{\st}$ and $\beta^{\st}$ remains accurate. We also point out the role played by the signal-to-noise ratio (SNR), where the noise is characterized by a clustered variance: the larger the SNR is, the larger the sample size has to be in order to accurately recover the clustering structure.\\
\indent Finally, we provide an application to income and democracy. We revisit the relationship between income and democracy, a central question in political economy studied by \cite{AcemougluEtAl2008}, \cite{BonhommeManresa2015}, and \cite{Zhang2023}. While our estimates of the regression parameters align with earlier findings, we do not detect evidence of multiple clusters in the data: the sample supports a single homogeneous group. This conclusion is robust across prior specifications. However, when we control for additional covariates, we detect a cluster structure with four groups. This means that the neglected controls have a strong signal compared to the variance of the clusters and so they blur the detection of the clustering structure.

The remainder of the paper is organized as follows. Section \ref{ss_literature} reviews related literature on panel data with group structures and on mixture of finite mixtures models. Section \ref{s:model} introduces the model, likelihood, and identification. Section \ref{ss:prior} describes the prior distribution. Section \ref{s:posterior} presents the posterior distribution and the Panel Data Telescoping Sampler. Section \ref{s:theoretical:validation} develops the asymptotic theory, with all the proofs collected in the Online Appendix. Section \ref{s:Monte:Carlo} reports results from Monte Carlo experiments. Section \ref{s:Application} contains the empirical application, and Section \ref{s:conclusions} concludes.

\subsection{Related literature}\label{ss_literature}
\indent Our paper connects to two strands of literatures: panel data models with group structures and Bayesian mixture models, particularly mixtures of finite mixtures (MFMs). In the first strand, a growing econometrics literature uses clustering methods to approximate heterogeneity in panel data. In this literature, clustering serves primarily as a dimension-reduction device: rather than modeling unit-level heterogeneity explicitly, researchers assume that individuals can be grouped into a finite number of types, each with its own parameters. This approach is particularly useful in short panels, where estimating a separate effect for each unit is difficult. Examples include \cite{BonhommeManresa2015,BonhommeLamadonManresa2022}, \cite{SuShiPhillips2016}, \cite{Zhang2023}. A key feature of this literature is that it does not assume the existence of a true clustering structure in the population. Instead, groups are introduced for tractability, and the group-specific unobservables are often allowed to vary over time. While this approach has proven highly influential, it differs fundamentally from ours. We develop a structural model of clustering, in which the population is assumed to be genuinely partitioned into a finite set of latent groups generated by a probabilistic mechanism. Our objective is not merely to approximate heterogeneity but to recover the underlying structure itself.\\
\indent Our paper also contributes to the Bayesian literature on MFMs. MFMs, introduced by \cite{Phillips1996bayesian} and \cite{RichardsonGreen1997} and further studied by \cite{Stephens2000}, \cite{Nobile2004}, \cite{NobileFearnside2007}, \cite{McCullaghYang2008}, \cite{GengBattacharyaPati2019},  \cite{XieXu2020}, \cite{FruwirthSchnatter2021} and \cite{GuhaHoNguyen2021} among others, provide a flexible prior over partitions by treating the number of mixture components as a random variable. While this literature has largely focused on \textit{i.i.d.} data and clustering of observable variables, we extend MFMs to a dynamic panel data setting with exogenous and predetermined covariates where the clustering structure concerns latent variables. A central issue in this literature is posterior consistency of the mixing distribution. While early work (e.g.,\cite{GhosalVanderVaart2001}) has focused on posterior consistency of the mixture distribution, more recent contributions such as \cite{Nguyen2013}, \cite{Scricciolo2019}, and \cite{OhnLin2023} established the posterior consistency of the mixing distribution in MFM models in \textit{i.i.d.} settings with no covariates. Our results complement this line by showing posterior contraction of the latent mixing distribution in panel settings, measured in Wasserstein distance, with implications for the recovery of both clusters and regression parameters.\\
\indent A widely used alternative for modeling clustering is the Dirichlet Process Mixture (DPM). However, DPMs are known to be inconsistent in recovering the cluster structure when the true number of clusters is finite: the common practice of making inference on $K$ via the DPM, simply by looking at the number of support points in the Dirichlet's posterior sample, makes the number of estimated clusters to grow with sample size, leading to spurious over-partitioning (\cite{MillerHarrison2013}). Recent work by \cite{Alamichel2024} extends these inconsistency results to the Pitman-Yor process mixture models, Gibbs-type processes and finite-dimensional representations of it (including the Dirichlet multinomial process and the normalized generalized gamma multinomial processes). Thus, the idea that a consistent estimate of the mixture distribution may lead to a consistent estimate of the number of mixture components and of the clusters is not correct, see \textit{e.g.} \cite{Leroux1992}. While some remedies exist -- \textit{e.g.}, \cite{AscolaniLijoiRebaudoZanella2022} show that consistency can be restored under specific priors on the concentration parameter -- our approach avoids these issues by directly modeling the number of clusters as finite but unknown.\\
\indent To summarize, our paper bridges the gap between the econometrics literature on panel clustering—which uses groups as an approximation tool without modeling their structural origin—and the Bayesian literature on MFMs, which provides a principled framework for inference on finite partitions but has not been adapted to panel settings and latent variables. By combining these perspectives, we provide both a structural interpretation of clustering in panel data and a computationally efficient algorithm for inference, supported by theoretical guarantees.

\paragraph{Notation.} We introduce here part of the notation used in the paper. Additional notations will be introduced later on in the manuscript and in the Online Appendix. For every integer $M\in\mathbb{N}$, we use the notation $[M]:=\{1,\ldots,M\}$. The empirical mean over cross-section units is written as $\EE_N[\cdot]:=\frac{1}{N}\sum_{i=1}^N [\cdot]$. For two conditional densities $f_1(y|z)$, $f_2(y|z)$ we denote the $L^1$-distance as $\|f_1(\cdot|z) - f_{2}(\cdot|z)\|_1 := \int|f_1(y|z) - f_{2}(y|z)|dy$ and the squared Hellinger distance as $h^2(f_1(\cdot|z),f_{2}(\cdot|z)) := \int(\sqrt{f_1(y|z)} - \sqrt{f_{2}(y|z)})^2 dy$. The Kullback-Leibler (KL) divergence between $f_1(y|z)$ and $f_2(y|z)$ is denoted by $\mathcal{KL}(f_1(\cdot|z)||f_2(\cdot|z)) := \bigintsss \log\left(\frac{f_1(y|z)}{f_2(y|z)}\right) f_1(y|z)dy$ and the KL second moment by $\mathcal{KL}_2(f_1(\cdot|z)||f_2(\cdot|z)) := \bigintsss \left(\log\left(\frac{f_1(y|z)}{f_2(y|z)}\right)\right)^2 f_1(y|z)dy.$\\
\indent For a set $\mathcal{T}$, a metric $\rho$, and a $\varepsilon>0$, we denote by $D(\varepsilon,\mathcal{T},\rho)$ the $\varepsilon$-packing number of $(\mathcal{T},\rho)$, that is, the maximum number of points that are mutually separated by at least $\varepsilon$ in distance. It is related to the covering number $N(\varepsilon,\mathcal{T},\rho)$ of $(\mathcal{T},\rho)$ by $N(\varepsilon,\mathcal{T},\rho) \leq D(\varepsilon,\mathcal{T},\rho) \leq N(\varepsilon/2,\mathcal{T},\rho)$. The symbols $\asymp$, $\lesssim$ and $\gtrsim$ denote equality and inequalities up to a constant.

\section{The model}\label{s:model}
  Let $\{y_{i,t}\}$ and $\{z_{i,t}\}$ be a univariate and a $p$-dimensional stochastic processes, respectively. Both $\{y_{i,t}\}$ and $\{z_{i,t}\}$ are strictly stationary, ergodic and observable. In addition, we take into account latent heterogeneity random variables $\{\alpha_i,\sigma_i^2\}$ and $\{u_{i,t}\}$, the first capturing the individual $i$'s specific heterogeneity and the second one capturing heterogeneity specific to individual and time. We consider the following panel data model: for every $i=1,\ldots,N$, $t=1,\ldots,T$, and every $h\geq 0$,
  \begin{multline}
    y_{i,t} = \gamma y_{i,t-1} + \beta'z_{i,t-h} + \alpha_i + u_{it},\qquad \\
    u_{i,t}| \{y_{i,s-1}\}_{s\in[t]},\{z_{i,s}\}_{s\in[T]},\alpha_i,\sigma_i^2 \sim \mathcal{N}(0,\sigma_i^2), \label{eq:2}
  \end{multline}
  \noindent where $|\gamma| < 1$, $\EE[u_{i,t}u_{j,t}] = 0$ for every $i\neq j$, and $\EE[u_{i,t}u_{i,t'}] = 0$ for every $t\neq t'$. The exogenous covariates $z_{i,t-h}\in\mathbb{R}^p$ and the predetermined covariate $y_{i,t-1}$ have homogeneous effects on the outcome $y_{i,t}$ captured by the vector of common parameters $(\gamma,\beta')'\in (-1,1)\times \mathbb{R}^p$. For simplicity, we consider only one lagged value of $y_{i,t}$. From \eqref{eq:2}, it follows that $\EE[u_{i,t} \alpha_i] = \EE[u_{i,t} y_{i,s-1}] = \EE[u_{i,t} z_{i,\tau}] = 0$ for all $i\in[N]$, $t,\tau \in [T]$ and $s\in [t]$, and that all the serial correlation in $y_{i,t}$ is captured by $y_{i,t-1}$ and $z_{i,t-h}$.  
  Under the assumption of Gaussianity of $u_{it}$, the conditional distribution of the outcome is Gaussian: $y_{it}|\gamma, \beta, \alpha_i, \sigma_i^2, x_{i,t-h}, y_{i,t-1} \sim \mathcal{N}(\gamma y_{i,t-1} + \beta'z_{i,t-h} + \alpha_i,\sigma_i^2)$.\\
  \indent We interpret the $(\alpha_i,\sigma_i^2)$ as unobservable random variables that are generated from the following finite mixing distribution independently on $z_{i,t}$ for every $t$: for every $\alpha_i\in\mathbb{R}$, $\sigma_i^2\in\mathbb{R}_+$,
  \begin{equation}\label{eq:finite:mixing}
    m \equiv m(\alpha_i,\sigma_i^2|K,\{\theta_j,w_j\}_{j\in[K]}) := \sum_{j=1}^{K}w_j\delta_{\theta_j}(\alpha_i,\sigma_i^2),
  \end{equation}
  where $\bw_K:=(w_1,\ldots, w_K) \in \Delta_{K} := \left\{(w_1,\ldots,w_{K})\in[0,1]^{K}; \,\sum_{j=1}^{K}w_j = 1\right\}$, $\theta_{j} := (\theta_{j,\alpha},\theta_{j,\sigma^2})'\in\mathbb{R}\times\mathbb{R}_+$ for $j\in[K]$, and $\btheta_K := (\theta_1, \ldots, \theta_{K})'\in\mathbb{R}^{K}\times\mathbb{R}_+^K$ is the matrix of $K$ support points of the distribution $m$. The $\theta_j$'s are the $K$ distinct values that the individual heterogeneities $\{(\alpha_i,\sigma_i^2)\}_{i\in [N]}$ can take on. 
  If we constraint each of the support points $\theta_j$ to belong to a compact set $\Theta := [-L,L]\times[\underline{\sigma}^2,\ol{\sigma}^2]\subset \mathbb{R} \times (0,\infty)$ for fixed values $0<\underline{\sigma}^2<\ol{\sigma}^2 < \infty$ and $L>0$, then, the distribution $m$ is an element of the set of $K$ atomic distributions with bounded support defined as:
  $$\mathcal{M}_{\leq K}(\Theta) := \left\{\sum_{j=1}^{K}w_j\delta_{\theta_j}; \, (w_1,\ldots,w_{K})\in\Delta_{K},\, (\theta_1,\ldots,\theta_{K})\in\Theta^{K}\right\}.$$
  \indent In the following, we denote by $\zeta := \{\gamma,\beta,\btheta_K,\bw_K,K\}$ the array collecting all the parameters of the model and denote $\theta_{j,\sigma} := \sqrt{\theta_{j,\sigma^2}}$ the $j$-th value of the standard deviation of the error term $u_{it}$. \\
  \indent By introducing for each observation $i\in[N]$ a latent allocation variable $\chi_i$ that assigns individual $i$ to component $j\in[K]$ with probability $w_j$, we can write model \eqref{eq:2}-\eqref{eq:finite:mixing} as a hierarchical latent variable model:
  \begin{align}
    & \chi_i| K, \bw_K \sim MulNom(1;w_1,\ldots,w_K),\quad \textrm{independently for }i\in[N], \label{eq:distribution:chi}\\
    & y_{it}|y_{i,t-1}, z_{i,t-h}, \chi_i = k,\beta, \gamma,\theta_k, K \sim \mathcal{N}(\gamma y_{i,t-1} + \beta'z_{i,t-h} + \theta_{k,\alpha},\theta_{k,\sigma^2}), \label{eq:distribution:conditional:chi}
  \end{align}
  \noindent where $MulNom$ denotes the multinomial distribution with only one number of trials and with $Prob(\chi_i = j|K,\bw_K) = w_j$ for every $j\in[K]$. The outcome of this multinomial distribution can be seen as a $K$-vector with one element equal to $1$ and all other elements equal to $0$. It is entirely controlled by the probabilities in $\bw_K$, where for every $k\in[K]$, $w_k$ is the probability that the $i$-th individual belongs to group $k$. This writing of the model will appear useful to draw from the posterior distribution.
%
    %
  \subsection{The likelihood.}\label{ss:likelihood} 
  Let $\by_i :=(y_{i1},\ldots,y_{iT})'$ be the $T$-vector of observations for the $i$-th unit, $\by := (\by_1,\ldots,\by_N)$ be a $(T\times N)$-matrix, $\by_0 := (y_{1,0},\ldots, y_{N,0})'$ be the $N$-vector of initial conditions, $\bz_{i} := (z_{i,1-h},\ldots, z_{i,T-h})'$ be the $(T\times p)$ matrix of strictly exogenous covariates, and $\bZ := (\bz_{1},\ldots,\bz_N)$ be the $T\times Np$ matrix of strictly exogenous covariates. We consider the conditional likelihood of the model given $\{\bz_i,y_{i,0}\}_{i\in[N]}$. 
  Conditional on the latent time-invariant allocation variable $\chi_i$, the joint distribution of $\by_i$ given $(\bz_i,y_{i0},\{\chi_i = k\}, \gamma, \beta, \theta_k, K)$ writes as
  \begin{displaymath}
    \by_{i}|\bz_{i}, y_{i0}, \chi_i = k, \gamma, \beta, \theta_k, K \sim \prod_{t=1}^T \phi\left(\frac{y_{it}-\gamma y_{i,t-1} - \beta'z_{i,t-h} - \theta_{k,\alpha}}{\theta_{k,\sigma}}\right)\frac{1}{\theta_{k,\sigma}},
  \end{displaymath}
  \noindent where $\phi(y)$ denotes the univariate density function of a $\mathcal{N}(0,1)$ distribution evaluated at $y$. Instead of conditioning on the latent allocation variable $\chi_i$, one can integrate out $(\alpha_i,\sigma_i^2)$ from the joint distribution of $\by_{i}|\bz_i, y_{i0}, \gamma, \beta, \alpha_i, \sigma_i^2$ with respect to $m(\cdot,\cdot|K,\btheta_K,\bw_K)$. By doing so, we get a joint distribution $P_{i,m,0}$ conditional on $(\bz_{i},y_{i,0},\zeta)$ whose Lebesgue density evaluated at $\by_{i}$ is
  \begin{multline}\label{eq:3}
    f_{\zeta}(\by_{i}|\bz_{i},y_{i0}) \equiv f(\by_{i}|\bz_{i}, y_{i0}, \zeta)\\
    := \bigintsss \prod_{t=1}^T\phi\left(\frac{y_{it}-\gamma y_{i,t-1} - \beta'z_{i,t-h} - \alpha_i}{\sigma_i}\right)\frac{1}{\sigma_i} m(d\alpha_i,d\sigma_i^2|K,\btheta_K,\bw_K)\\
    = \sum_{j=1}^{K}w_j \prod_{t=1}^T \phi\left(\frac{y_{it}-\gamma y_{i,t-1} - \beta'z_{i,t-h} - \theta_{j,\alpha}}{\theta_{j,\sigma}}\right)\frac{1}{\theta_{j,\sigma}}.
  \end{multline}
  \noindent The joint conditional likelihood of the model, denoted by $\ell(\zeta|\by;\bZ,\by_0)$ writes as:
  \begin{equation}
    \ell(\zeta|\by;\bZ,\by_0) := \prod_{i=1}^N f_{\zeta}(\by_{i}|\bz_{i},y_{i0}). 
  \end{equation}
  The corresponding conditional distribution of the whole sample, given $\bZ,\by_0,\zeta$ is denoted by $P_{m,0}^{(N)} := \bigotimes_{i=1}^N P_{i,m,0}$.

  \indent \paragraph{Remark 1.} \textit{The joint likelihood function can be written in an alternative way by making explicit the partitions of the $N$ individuals $\{1,\ldots,N\}$ into $K$ groups. To this purpose, we use the latent allocation variable $\chi_i$ in \eqref{eq:distribution:chi} that assigns a group to individual $i$ and we introduce the set $E_k :=\{i\in[N]; \chi_i=k\}$, for every $k\in[K]$. Moreover, every sequence of sets $\{E_k\}_{k\in[K]}$ such that $E_k\cap E_{k'} = \emptyset$, $\forall k\neq k'$, and $\bigcup_{k\in[K]} E_k = [N]$ defines a partition $\mathcal{C}_K$ of the set $[N]$ into $K$ groups and we denote by $\mathfrak{C}_K$ the set of all the partitions of $[N]$ into $K$ groups, so that $\mathcal{C}_K\in\mathfrak{C}_K$. The set $\mathfrak{C}_K$ has $K^N$ elements. With this notation, and by using the hierarchical latent variable model \eqref{eq:distribution:chi}-\eqref{eq:distribution:conditional:chi} we can write:}
  \begin{multline*}
    \ell(\zeta|\by;\bZ,\by_0) 
    = \sum_{\mathcal{C}_K\in\mathfrak{C}_K}\prod_{j=1}^K \left\{\prod_{i\in E_j} w_j \prod_{t=1}^T \phi\left(\frac{y_{it}-\gamma y_{i,t-1} - \beta'z_{i,t-h} - \theta_{j,\alpha}}{\theta_{j,\sigma}}\right)\frac{1}{\theta_{j,\sigma}}\right\}\\
    = \sum_{\mathcal{C}_K\in\mathfrak{C}_K}w_1 ^{n_1}\cdot \ldots \cdot w_K ^{n_K}\prod_{j=1}^K \prod_{i\in E_j} \prod_{t=1}^T \phi\left(\frac{y_{it}-\gamma y_{i,t-1} - \beta'z_{i,t-h} - \theta_{j,\alpha}}{\theta_{j,\sigma}}\right)\frac{1}{\theta_{j,\sigma}},
  \end{multline*}
  \noindent \textit{where $n_j = |E_j|$, $\forall j\in[K]$, and $\sum_{j\in[K]}n_j = N$.}\\

 \paragraph{True sampling distribution.} The true sampling distribution of the $T$-random vector $\by_i$ conditional on $(\bz_{i}, y_{i0})$ is denoted by $P_{i,0}^{\st}$, has Lebesgue density $f_{\zeta^{\st}}(\cdot|\bz_{i}, y_{i0})$ and takes the form of \eqref{eq:3} with $\zeta$ replaced by its true value $\zeta^{\st} := \{\gamma^{\st}, \beta^{\st}, \btheta^{\st}, \bw^{\st},K^{\st}\}$ where we use the simplified notation $\btheta^{\st} \equiv \btheta_{K^{\st}}^{\st}$ and $\bw^{\st} \equiv \bw_{K^{\st}}^{\st}$. It is a mixture with respect to the $K^{\st}$-atomic distribution $m^{\st} \equiv m^{\st}(\cdot,\cdot|K^{\st},\btheta^{\st},\bw^{\st})$ of $(\alpha_i,\sigma_i^2)$, where $m^{\star}\in\mathcal{M}_{\leq K^{\star}}(\Theta)$ and $K^{\st}\in\mathbb{N}$ is the true number of components in the mixture. The true conditional distribution of the whole sample, given $\bZ,\by_0,\zeta^{\st}$ is denoted by $P_{0}^{\st(N)} := \bigotimes_{i=1}^N P_{i,0}^{\star}$ and the expectation taken with respect to $P_{0}^{\st (N)}$ is denoted by $\mathbf{E}^{\st}[\cdot]$.\\
 \indent Since $K^{\st}$ is supposed to be unknown and is allowed to take any value in $\mathbb{N}_+$, then assuming that the true $P_{i,m}^\star$ has a Lebesgue density of the form \eqref{eq:3} is not restrictive. Indeed, any distribution can be well approximated by a Gaussian mixture with a potentially infinite number of components. Therefore, a potential misspecification error is very small here.\\
 \indent To guarantee identification, we assume in the following that $w_{j}^{\st} >0$ for every $j\in[K^{\st}]$ and that for every $j\neq k$ either $\theta_{j,\alpha}^{\st} \neq \theta_{k,\alpha}$ or $\theta_{j,\sigma^2}^{\st} \neq \theta_{k,\sigma^2}$ or both. 
Therefore, $K^{\st}$ is defined as the true number of components in the mixture with nonzero weights and with corresponding parameters that differ in at least one of the two dimensions, that is,
  $$K^* := \sharp\{k; w_k^* > 0 \textrm{ and } \forall j\neq k, \theta_{j,l}^{\st} \neq \theta_{k,l}^{\st} \textrm{ for at least one } l\in \{\alpha,\sigma^2\} \,\}.$$
\indent When a sample of size $N$ is observed, which is a realization of a draw from the true model, it might be that realizations from only some of the $K^{\st}$ components are observed. We denote by $K_{+}^{\st} \equiv K_{+,N}^{\st}$ the number of the mixture components that have realized and we call them the realized components given $N$. The number $K_+^{\st}$ increases with $N$ and converges to $K^{\st}$ as $N\rightarrow \infty$. 
We denote by $\bw_+^{\st} \equiv \bw_{+,N}^{\st}$ the associated $K_+^{\st}$-vector of true mixing probabilities, conditional on $N$. Each component of $\bw_+^{\st}$ equals the corresponding component of $\bw^{\st}$ normalized so that $\sum_{j=1}^{K_+^{\st}} w_{+,j}^{\st} = 1$, where $\bw_+^{\st} := (w_{+,1}^{\st},\ldots, w_{+,K_+^{\st}}^{\st})'$. That is, $w_{+,j}^{\st} = w_j^{\st}/\sum_{j=1}^{K_+^{\st}}w_{+,j}^{\st}$ for every $j\in[K_+^{\st}]$. Therefore, conditional on $N$, we have
  $$m^{\st}(\alpha_i,\sigma_i^2|N,K_+^{\st},\btheta^{\st},\bw_+^{\st}) := \sum_{j=1}^{K_+^{\st}}w_{+,j}^{\st}\delta_{\theta_{j}^{\st}}(\alpha_i,\sigma_i^2).$$
\subsection{Identification.}
In this section we look at the identification of the structural mechanism, which is fully characterized by the parameters $\zeta^{\st}$.
  \begin{defn}\label{def_identifiability_conditionaò}
    We say that the mixture model \eqref{eq:3} is identified if
    $$f_{\zeta_1}(\by_i|\bz_{i},y_{i0}) = f_{\zeta_2}(\by_i|\bz_{i},y_{i0}),$$
    \noindent where $\zeta_{\ell} := (\gamma_{\ell},\beta_{\ell},\btheta_{K_{\ell},\ell},\bw_{K_{\ell},\ell},K_{\ell})$ for $\ell = 1,2$, if and only if $\gamma_1 = \gamma_2$, $\beta_1=\beta_2$, $K_1 = K_2$ and the components in the sums can be ordered so that $w_{1,j} = w_{2,j}$ and $\theta_{1,j} = \theta_{2,j}$, for all $j\in[K_1]$.
  \end{defn}
  \indent We denote by $\Phi_T(\mathcal{y};a_1,a_2)$ the cumulative distribution function of a $T$-dimensional Gaussian distribution with mean $a_1$ and variance $a_2$ evaluated at $\mathcal{y}\in\mathbb{R}^{T}$, and by $\phi_T(\mathcal{y};a_1,a_2)$ its Lebesgue density evaluated at $\mathcal{y}$. 
  Let us consider the class of $T$-dimensional conditional Gaussian cumulative distribution functions (cdf's), given $\bz_{i}$, $y_{i0}$, $\beta \in\mathbb{R}^p$, and $\gamma \in (-1,1)$, with mean $\mu_{1:T}^0(\theta_{\alpha},\beta,\gamma,y_{i0},\bz_i)$ and variance-covariance matrix $\frac{\theta_{\sigma^2}}{1 - \gamma^2} V_{T}^0$:
    \begin{displaymath}
      \mathcal{F}(\bz_{i},y_{i0},\gamma,\beta) := \Big\{\Phi_T\left(\mathcal{y};\mu_{1:T}^0(\theta_{\alpha},\gamma,\beta,y_{i0},\bz_i), \frac{\theta_{\sigma^2}}{1 - \gamma^2} V_{T}^0\right), \, \mathcal{y}\in\mathbb{R}^T, \,\theta_{\alpha}\in\mathbb{R}, \, \theta_{\sigma^2}\in\mathbb{R}_+ \Big\},
    \end{displaymath}
    where the $T$-vector $\mu_{1:T}^0(\theta_{\alpha},\gamma,\beta,y_{i0},\bz_i)$ and the $T$-symmetric matrix $V_T^0$ are defined in the Supplementary Material G and $V_T^0$ is a deterministic function of $\gamma$. 
    Let
    \begin{multline*}
      \mathcal{H}(\bz_{i}, y_{i0}) := \Big\{H(\cdot|\bz_i, y_{i0}); H(\cdot|\bz_i, y_{i0}) = \sum_{j=1}^K w_j \Phi_T(\cdot), \, w_j >0, \sum_{j=1}^K w_j = 1, \\
      \Phi_T(\cdot) \in \mathcal{F}(\bz_{i},y_{i0},\gamma,\beta), \,\beta\in\mathbb{R}^p,\, \gamma\in (-1,1),\,K=1,2,\ldots \Big\}
    \end{multline*}
    be the class of all finite mixtures of $\mathcal{F}(\bz_{i},y_{i0},\gamma,\beta)$. The following proposition guarantees identification of $\mathcal{H}(\bz_{i}, y_{i0})$ for every $\bz_{i}$, $y_{i0}$, and identification of $\zeta^{\st}$. Its proof is in Online Appendix A.1.

  \begin{prop}\label{prop:1:identification:conditional}
    Suppose that $\{y_{i,t}\}_t$ follows model \eqref{eq:2} with $|\gamma^{\st}| < 1$, then the class $\mathcal{H}(\bz_{i},y_{i0})$ is identifiable. Moreover, if the matrix $Var(z_{i,t})$ has full rank, then the parameters $\theta_{j,\alpha}^{\st},\theta_{j,\sigma^2}^{\st},\gamma^{\st}, \beta^{\st}$ are identifiable.
  \end{prop}

\section{Prior distribution}\label{ss:prior}
  In this section we describe the specification of the prior distribution for $\zeta$. 
  A prior on $(\btheta_K, \bw_K,K)$ induces a prior on the $K$-atomic distribution $m(\cdot,\cdot|K,\btheta_K,\bw_K)$. An important feature of clustering is that the prior for $(\btheta_K,\bw_K,K)$ has to be informative because, in a mixture setting, a non-informative prior might result in an improper posterior distribution if there are no observations allocated in some components. We use the same notation $\Pi$ for the marginal and the joint prior distribution as well as for their Lebesgue densities. Our prior specification is the following:
  \begin{eqnarray*}
    K|N & \sim & \Pi(K|N),\\
    \varphi, v & \sim & \Pi(\varphi)\Pi(v)\\
    \btheta_K|K,\varphi & \sim & \prod_{k=1}^K \Pi(\theta_{k,\alpha};\varphi_1)\Pi(\theta_{k,\sigma^2};\varphi_2),\\
    \bw_K|K,v & \sim & \Pi(\bw_K;K,v),\\
    (\gamma,\beta) & \sim & \Pi(\gamma;\varpi_1) \Pi(\beta;\varpi_2),
  \end{eqnarray*}
  \noindent where $\varpi := (\varpi_1,\varpi_2)$ is a fixed parameter, $\varphi := (\varphi_1,\varphi_2)$, $\Pi(\bw_K;K,v)$ has support $\Delta_K$ and $\Pi(\gamma;\varpi_1)$ has support $(-1,1)$. Conditional on $K$, the random vectors $\btheta_K$ and $\bw_K$ are independent. Model \eqref{eq:distribution:chi}-\eqref{eq:distribution:conditional:chi} together with the prior on $K,\btheta_K,\bw_K$ given above belongs to the class of mixture of finite mixtures (MFMs) (\textit{e.g.}, \cite{RichardsonGreen1997}, \cite{Nobile2004}, and \cite{MillerHarrison2018}). In this paper we extend the MFM to a panel data setting with predetermined regressors.\\
  \indent The prior distribution for $K$ can be any distribution with support $\{1,2,\ldots\}$ and it can depend on $N$ through its hyperparameters. Examples are: (1) the translated Binomial distribution where $K-1\sim \mathcal{B}in(K_{\max},p)$ for some $K_{\max}>1$ and $p\in[0,1]$, (2) the Poisson distribution: $K-1\sim \mathcal{P}oi(\lambda)$ for $\lambda > 0$, and (3) the geometric distribution: $K-1\sim Geometric(q)$ for $q\in[0,1]$. The motivation for making the prior of $K$ dependent on $N$ is to reproduce a kind of ascending clustering, that is as $N$ is small on can think that every individual forms a different clustering. As more observations arrive, one could prefer either to attribute them to existing groups (shrinking prior) or to create new groups (spreading our prior). Our asymptotic results require a prior that penalizes mixing distributions with too many components, see Assumption \ref{ass:1} \textit{(iii)}.\\
  \indent The prior of $\bw_K$ depends on a hyperparameter $v\in\mathbb{R}$. Depending on whether $v$ varies or not with $K$ we have a dynamic MFM: $v = e_0/K$, or a static MFM: $v=e_0$, for a given hyperparameter $e_0$. The hyperparameter $e_0$ can be fixed to a value or endowed with a prior distribution. An example of a prior for $\bw_K$ is the symmetric Dirichlet distribution of order $K$ where $\bw_K|K,v \sim \mathcal{D}ir(v,\ldots, v)$ with $v > 0$ the concentration parameter. This is the prior we use in our implementation. A symmetric Dirichlet distribution is well-suited if one does not want to favor a priori any component of the mixture over another.\\
  \indent Examples of priors for $\btheta_K$ are: (1) the multivariate uniform distribution on $[-L,L]^{K}\times[\underline{\sigma}^2,\ol{\sigma}^2]^K$: $\Pi(\btheta_K|K,\varphi) = \prod_{j=1}^K (2L)^{-1}(\ol{\sigma}^2 - \underline{\sigma}^2)^{-1}$; (2) the product of $K$ truncated Normal - inverse Gamma distributions truncated on the interval $[-L,L] \times [\underline{\sigma}^2,\ol{\sigma}^2]$.

\subsection{Prior on the number of clusters in the sample}
As already discussed in Section \ref{ss:likelihood}, it is useful to distinguish between the random parameter $K$, which is the number of components of the mixture model in the population, and the random parameter $K_{+,N}$, which is the number of non-empty components (or clusters) in the sample. The latter is the random parameter corresponding to the number of the mixture components from which the data have originated and is defined as $K_{+,N} := \sum_{k=1}^{K} 1\{N_k >0\}$, where $N_k := \sharp \{i\in[N];\chi_i =k\}$ for $k \in [K]$ are the cluster sizes. It is a deterministic function of the vector of latent allocation variables $\chi := (\chi_1,\ldots,\chi_N)$ and it is a non-decreasing function of the sample size $N$. If $\chi$ is known then $K_{+,N}$ is known. For brevity we write in the following $K_+ := K_{+,N}$. The prior $\Pi(K_{+} = k|N,K,v)$ for $K_{+}$, conditional on the number of components $K$, on $v$, and on the sample size $N$, can then be obtained from the prior probability mass function $\Pi(N_1,\ldots, N_k|N,K,v)$ of the labeled cluster sizes $(N_1,\ldots,N_k)$ of a partition with $k$ non-empty clusters such that $N_1 + \ldots + N_k = N$. The resulting prior is: for every $k\in [K]$,
\begin{equation}\label{eq:prior:Kplus}
  \Pi(K_{+} = k|N,K,v) = \sum_{\substack{N_1,\ldots,N_k >0;\\ N_1 + \ldots + N_k = N}}\Pi(N_1,\ldots,N_k|N,K,v).
\end{equation}
In the case where $\bw_K|K,v \sim\mathcal{D}ir(v)$, then
$$\Pi(N_1,\ldots,N_k|N,K,v) = \binom{K}{k} \binom{N}{N_1,N_2,\ldots, N_{k}} \frac{\Gamma(v K)}{\Gamma(v K + N)}\prod_{j = 1}^k\frac{\Gamma(N_j + v)}{\Gamma(v)},$$
\noindent where $\binom{K}{k}$ denotes the number of possible ways to choose $k$ non-empty clusters among the $K$ components and the multinomial coefficient $\binom{N}{N_1,N_2,\ldots, N_{k}}$ denotes the number of ways to assign $N$ observations into $k$ clusters of size $N_1,\ldots, N_{k}$. The last factor $\frac{\Gamma(v K)}{\Gamma(v K + N)}\prod_{j = 1}^k\frac{\Gamma(N_j + v)}{\Gamma(v)}$ accounts for the marginal probability distribution of the latent vector $\chi := (\chi_1,\ldots,\chi_N)$: $\Pi(\chi|K,v)$. 
\indent Finally, by integrating out $K$ from \eqref{eq:prior:Kplus} with respect to its prior distribution we get: for every $k \in [K]$,
\begin{equation}
  \Pi(K_{+} = k|N,v) = \sum_{K = k}^{+\infty} \Pi(K|N) \Pi(K_{+} = k|N,K,v).
\end{equation}
\noindent The induced prior $\Pi(K_{+} = k|N,v)$ for MFM models has been derived in \cite[Section 3.2]{FruwirthSchnatter2021} for various prior distributions on $K$. In Table \ref{tab:Mean:K:plus} we illustrate how the prior mean of $K_+$, given $(N,v)$, is affected by the sample size $N$, the hyperparameter $v$ of the prior $\Pi(\bw_K;K,v)= \mathcal{D}ir(v)$, and the hyperparameters of the prior for $K$, which is taken to be a translated Negative Binomial prior $NB(a,p)$ with $a >0$ and $p\in[0,1]$. For every values of $a$ and $p$ considered, the prior mean of $K$ is equal to $a+1$. We expect that as $N$ increases, the prior expectation of $K_+$ converges towards the prior expectation of $K$. For all the three values of $a$ we observe convergence and we notice that the prior of $K$ does not affect too much the convergence properties of the prior mean of $K_+$. Instead, the latter is much more sensitive to the choice of the hyperparameter $v$ of the Dirichlet prior for $\bw_K$. Figure \ref{fig:effect:N:dynamic} in the Appendix plots the posterior mean of $K_+$ as a function of the sample size $N$ for the static and dynamic MFM and for differentvalues of $v$. The dashed black line corresponds to the prior mean of $K$, while the three curves correspond to the prior mean of $K_+$ for three different priors for $K-1$: Geometric (green line), Poisson (blue) and Negative Binomial (red). We see that convergence is observed for $v=1$, while for $v<1$ and $v>1$ the prior mean of $K_+$ fails to converge to the prior mean of $K$.

\begin{table}[ht]
\centering
\footnotesize{
\begin{tabular}{|c|c||c|c|c||c|c|c||}
  \hline
  & & \multicolumn{3}{c}{$N=50$} & \multicolumn{3}{c}{$N=200$}\\
 \hline
 \hline
 & & {\tiny $NB(1,0.5)$} & {\tiny $NB(4,0.5)$} & {\tiny $NB(9,0.5)$} & {\tiny $NB(1,0.5)$} & {\tiny $NB(4,0.5)$} & {\tiny $NB(9,0.5)$} \\
 \hline
 \multirow{6}{*}{\rotatebox[origin=c]{90}{Static MFM}} & $v = e_0 = 0.5$ & 1.77 & 4.01 & 7.11 & 1.85 & 4.50 & 8.48\\
 & $v = e_0 = 1$ & 1.93 & 4.63 & 8.23 & 2.00 & 4.90 & 9.50\\
 & $v = e_0 = 6$ & 1.97 & 5.02 & 9.53 & 2.01 & 5.01 & 9.92\\
 \cline{2-8}
 & $v\sim\mathcal{G}a(1,0.5)$ & 1.18 & 1.79 & 2.63 & 2.06 & 5.00 & 9.81\\
 & $v\sim\mathcal{G}a(1,1)$ & 2.02 & 4.95 & 9.04 & 2.02 & 5.02 & 9.86\\
 & $v\sim\mathcal{G}a(8,1)$ & 2.03 & 5.04 & 9.65 & 1.97 & 5.00 & 9.96\\
 \hline
 \multirow{6}{*}{\rotatebox[origin=c]{90}{Dynamic MFM}} & $v=e_0 =0.5$ & 1.77 & 3.98 & 7.17 & 1.88 & 4.50 & 8.48 \\
 & $v=e_0 = 1$ & 1.96 & 4.61 & 8.18 & 1.98 & 4.90 & 9.51\\
 & $v=e_0 = 6$ & 2.00 & 5.02 & 9.52 & 2.02 & 5.01 & 9.92\\
 \cline{2-8}
 & $v\sim\mathcal{G}a(1,0.5)$ &1.39 & 2.45 & 3.97 & 2.00 & 4.95 & 9.65\\
 & $v\sim\mathcal{G}a(1,1)$ & 1.86 & 4.29 & 7.54 & 1.70 & 3.51 & 6.29\\
 & $v\sim\mathcal{G}a(8,1)$ & 2.01 & 5.02 & 9.58 & 2.03 & 5.02 & 9.97\\
  \hline
\end{tabular}
}
\caption{{\footnotesize Prior expectation of $K_+$, given $N\in\{50,200\}$ and $K$ drawn from $\Pi(K-1) = NB(a,p)$, when $\bw_K|K,v\sim \mathcal{D}ir(v)$ for the two cases \textit{static MFM} and \textit{dynamic MFM}. In $\mathcal{G}a(a,b)$ the parameter $a$ denotes the shape and $b$ denotes the scale. The prior mean of $K$ for the three priors considered is $2$, $5$, and $10$, respectively. }}\label{tab:Mean:K:plus}
\end{table}

\section{Posterior Distribution and the Telescoping sampling algorithm}\label{s:posterior}
The posterior distribution of $\zeta$ is proportional to (by removing the hyperparameters to lighten the notation):
\begin{multline*}
\Pi (\zeta|\by,\bZ,\by_{0}) \propto  \Pi(K)\Pi(\bw_K|K)\Pi(\btheta_K|K)\Pi(\gamma,\beta)\times\\
\sum_{\mathcal{C}_{K_+}\in\mathfrak{C}_K}w_1 ^{n_1}\cdot \ldots \cdot w_K ^{n_K}\prod_{j=1}^K \prod_{i\in E_j} \prod_{t=1}^T \phi\left(\frac{y_{it}-\gamma y_{i,t-1} - \beta'z_{i,t-h} - \theta_{j,\alpha}}{\theta_{j,\sigma}}\right)\frac{1}{\theta_{j,\sigma}},
\end{multline*}
\noindent where $\mathcal{C}_{K_+}$ denotes the partition of $[N]$ into $K_+$ clusters. More precisely, the partition $\mathcal{C}_{K_+}$ writes $\mathcal{C}_{K_+} = \{E_1,\ldots,E_{K_+}\}$, where each cluster $E_k$ contains all the observations generated by the same mixture component, that is, $E_k := \{i\in[N];\chi_i = k\}$ for every $k\in[K]$.\\
%
%
\indent To draw from the posterior distribution of a MFM one can use the Reversible Jump MCMC of \cite{RichardsonGreen1997}. However, it has been shown (\textit{e.g.} Dellaportas \& Papageorgiou, 2006) that this sampler is challenging to tune in multidimensional cases. Another algorithm has been proposed in \cite{MillerHarrison2018}. Here, we propose to use the telescoping sampler of \cite{FruwirthSchnatter2021} and we extend it to a panel data regression model with predetermined and exogenous regressors. This is a trans-dimensional Gibbs sampler. Details of the sampler are provided in the Algorithm \ref{algorithm:1} below. The differences with respect to the original telescoping sampler of \cite{FruwirthSchnatter2021} is the introduction of the temporal dimension, which makes $\by$ in the algorithm to be a matrix, and of step (2)-(c) which takes into account the covariates (exogenous and predetermined).\\
\indent The idea of the telescoping sampler is that, instead of working with the marginal exchangeable partition probability function (EPPF) $\pi(\mathcal{C}_{K_+}|N,v)$ of the partition $\mathcal{C}_{K_+}$, as in \cite{MillerHarrison2018}, it works with the conditional EPPF $\pi(\mathcal{C}_{K_+}|N,K,v)$ by including $K$ as an additional latent variable, in addition to $\mathcal{C}_{K_+}$, in the sampling algorithm. The explicit inclusion of $K$ in the sampling algorithm is also present in \cite{RichardsonGreen1997}. However, instead of using the Reversible Jump MCMC scheme as in \cite{RichardsonGreen1997}, $K$ is sampled conditional on $\mathcal{C}_{K_+}$ from the conditional posterior $\pi(K|\mathcal{C}_{K_+},N,v) \propto \pi(K|N) \pi(\mathcal{C}_K|N,K,v)$. The latter is very convenient. Indeed, due to the conditional independence of $\theta_k$, $k\in[K]$, in the non-empty components and $K$, given the partition $\mathcal{C}_{K_+}$, $K$ is sampled from the conditional posterior $\pi(K|\mathcal{C}_{K_+},N,v)$ which does not depend on $\theta_k$. This makes the Telescoping Sampler easy to implement.\\
\indent The telescoping sampling samples $K$ and $K_+$, and the number of empty components $K-K_+$, which can be larger than or equal to zero, varies over the iterations of the sampler. As explained in \cite{FruwirthSchnatter2021}, the difference between $K$ and $K_+$, which can extend or contract to zero, behaves like a telescope and so, it gives the name to the sampler. In the algorithm, the hyperparameter $v$ of the prior on $\bw_K$ is endowed with a prior.

  \begin{algorithm}[!ht]
  \caption{telescoping sampler for dynamic panel data}
\footnotesize{
\KwData{$\by,\by_0,\bZ$.}
\KwInput{$\gamma,\beta,\btheta_K, \bw_K, K,\varphi$}
\begin{itemize}
  \item[(1)] Update the partition $\mathcal{C}_{K_+}$ by sampling from $\pi(\chi)$, where $\chi := (\chi_1,\ldots,\chi_N)'$:
  \begin{itemize}
    \item[(a)] sample $\chi_i$, for $i=1,\ldots,N$, from $\Pi(\chi_i = k|\by, \bZ,\by_0,K,w_k,\gamma,\beta,\theta_k)$;
    \item[(b)] determine $N_k := \sharp\{i; \chi_i = k\}$ for $k=1,\ldots,K$, and the number $K_+:= \sum_{k=1}^K\1\{N_k > 0\}$ of non-empty components and relabel such that the first $K_+$ components are non-empty.
  \end{itemize}
  \item[(2)] Conditional on $\mathcal{C}_{K_+}$, update the parameters of the non-empty components:
  \begin{itemize}
    \item[(a)] For the filled components $k=1,\ldots, K_+$ sample $\theta_k|\chi,\by,\bZ,\by_0,\gamma,\beta,\varphi$ from
    $$\Pi(\theta_k|\chi,\by,\bZ,\by_0,\gamma,\beta,\varphi) \propto \Pi(\theta_k|\varphi)\prod_{\{i;\chi_i = k\}} f(\by_i|\bz_i,y_{i0},\chi_i=k,\gamma,\beta,\theta_k).$$
    \item[(b)] (Optional) If a prior $\Pi(\varphi)$ on $\varphi$ is specified, then sample the hyperparameters $\varphi$ conditional on $K_+$ and $\btheta_{K_+}$ from
    $$\Pi(\varphi|\btheta_{K_+}, K_+) \propto \Pi(\varphi) \prod_{k=1}^{K_+}\Pi(\theta_k|\varphi).$$
    \item[(c)] Sample $(\gamma,\beta)$ from $\Pi(\gamma,\beta|\by,\bZ,\by_0,\chi,\btheta_{K_+},K_+,\varpi)$.
  \end{itemize}
  \item[(3)] Conditional on $\mathcal{C}_{K_+}$, draw new values of $K$ and $v$:
  \begin{itemize}
    \item[(a)] Sample $K$ from
    $$\Pi(K|\mathcal{C}_{K_+},N,v) \propto \Pi(K|N) \Pi(\mathcal{C}_{K_+}|N,K,v).$$
    \item[(b)] Use a random-walk Metropolis-Hastings step with proposal: $\log(v^{new})\sim \mathcal{N}(\log(v^{old}),s_{v^{old}}^2)$ to sample $v$ from: $\Pi(v|\mathcal{C}_{K_+},K) \propto \Pi(\mathcal{C}_{K_+}|K,v) \Pi(v)$.
  \end{itemize}
  \item[(4)] Conditional on $\chi,\varphi,K,v$, add $K - K_+$ empty components and update $\bw_K$:
  \begin{itemize}
    \item[(a)] If $K>K_+$, then add $K - K_+$ empty components (\textit{i.e.} $N_k = 0$ for $k= K_+ + 1,\ldots,K$) and sample $\btheta_k$ from the prior $\Pi(\btheta_k|K,\varphi)$ for $k= K_+ + 1,\ldots,K$.
    \item[(b)] Sample $\bw_K|K,v,\chi \sim\mathcal{D}ir(v + N_1,\ldots,v + N_K)$.
  \end{itemize}
  \item[(5)] Evaluate the Mixture Likelihood $\prod_{i=1}^N f_{\zeta}(\by_i|\bz_i,y_{i0})$.
\end{itemize}

\KwResult{$\{\gamma^{(j)},\beta^{(j)},K^{(j)}, K_+^{(j)}, \btheta_{K^{(j)}}^{(j)},\bw_{K^{(j)}}^{(j)}\}_{j\in [MC]}$.}
}
\label{algorithm:1}
\end{algorithm}

\section{Theoretical validation}\label{s:theoretical:validation}
This section studies the asymptotic behaviour of our posterior distribution for $N\rightarrow \infty$. It is divided in three parts. First, we state an assumption about the prior and show the posterior does not overestimate $K$. Then, in Section \ref{ss:consistency:static} we establish posterior consistency in the static case, that is, the panel data model \eqref{eq:2} without the dynamic component. Finally, in Section \ref{ss:consistency:dynamic} we extend this result to the dynamic model \eqref{eq:2} with the lagged dependent variable. Our results do not require $T$ to increase to infinity and it if kept fixed.
\subsection{Assumptions and preliminary results.}
The following assumptions concerns the prior distribution. According to it, the prior must place enough mass near the truth and penalize overly large values of $K$ as $N$ grows.
\begin{assum}\label{ass:1}
  (i) For any $K\in\mathbb{N}$ and any $(w_1^{\st},\ldots,w_{K}^{\st})\in\Delta_K$ there is a positive constant $c_0$ such that for any $\epsilon \leq \frac{1}{2}(1 - e^{-1})^2$,
  $$\Pi\left(\left.\sum_{j=1}^K |w_j^{\st} - w_j| \leq \epsilon\right|K,v\right) \gtrsim \epsilon^{c_0}.$$
  (ii) For any $K\in\mathbb{N}$ and any $\btheta^{\st}\in [-L,L]^K \times [\underline{\sigma},\ol{\sigma}]^K$, there exists a positive constant $c_1$ such that for any $\epsilon > 0$,
  $$\Pi\left(\max_{1\leq j\leq K}|\theta_{j,\alpha} - \theta_{j,\alpha}^{\st}|\leq \epsilon,\, \max_{1\leq j\leq K}|\theta_{j,\sigma^2} - \theta_{j,\sigma^2}^{\st}|\leq \epsilon|K, \varphi\right) \gtrsim \epsilon^{c_1}.$$
  (iii) The prior distribution on the number of components $K$ depends on $N$. There are a constant $c_3>0$ and a constant $A>0$ such that for any $N\in\mathbb{N}$ and any $k\in\mathbb{N}$,
  \begin{equation}
    \frac{\Pi\left(K = k + 1|N\right)}{\Pi\left(K = k|N\right)} \leq c_3e^{-A \log(N)}.
  \end{equation}
  (iv) The prior distribution on $\beta$ is such that: $\forall \eta >0$, $\forall \bz\in\mathbb{R}^{T\times p}$, and $\forall \beta^{\st}\in\mathbb{R}^p$,
    $$\Pi(\|\bz(\beta - \beta^{\st})\|_{\ell_1} \leq \eta | \bz,\varpi_2) \geq \eta^p T^{-p}.$$
  (v) The prior distribution on $\gamma$ is such that: there is a positive $c_2$ for which $\forall \epsilon > 0$ and $\forall \gamma^{\st}\in(-1,1)$,
    $$\Pi(|\gamma - \gamma^{\st}| \leq \epsilon|\varpi_1) \gtrsim \epsilon^{c_2}.$$
\end{assum}
The following prior distributions satisfy Assumption \ref{ass:1} \textit{(iii)} if the hyperparameters are chosen in an appropriate way:
  \begin{enumerate}
    \item Translated Binomial distribution where $K-1|N\sim \mathcal{B}in(K_{\max},p)$, for some $K_{\max}\in\mathbb{N}$ and $p\asymp N^{-A}$. 
    Assumption \ref{ass:1} \textit{(iii)} is satisfied because $\frac{\Pi\left(K = k + 1|N\right)}{\Pi\left(K = k|N\right)} = \frac{(K_{\max} - k + 1)p}{k(1-p)} \lesssim N^{-A}$ by using the inequality $(1 - p)\gtrsim 1$. 
    \item Negative Binomial distribution where $K-1|N\sim \mathcal{NB}(r,p)$ for some $r>0$ and $p\gtrsim 1 - N^{-A}$.
    \item Poisson distribution: $K-1|N\sim \mathcal{P}oi(\lambda)$ with $\lambda \asymp N^{-A}$. Assumption \ref{ass:1} \textit{(iii)} is satisfied because $\frac{\Pi\left(K = k + 1\right)}{\Pi\left(K = k\right)} = \frac{\lambda}{k} \lesssim N^{-A}$ for every $k\in\mathbb{N}$.
    \item Geometric distribution: $K-1|N\sim Geometric(q)$ with $q \gtrsim 1 - N^{-A}$. Assumption \ref{ass:1} \textit{(iii)} is satisfied because $\frac{\Pi\left(K = k + 1|N\right)}{\Pi\left(K = k|N\right)} = \frac{(1-q)^{k}q}{(1-q)^{k-1}q} = (1-q) \lesssim N^{-A}$.
  \end{enumerate}
In addition, a symmetric Dirichlet prior for $\bw_K$ with hyperparameter $v$, as discussed in Section \ref{ss:prior}, satisfies Assumption \ref{ass:1} \textit{(i)} for $v \in (0,1]$, see \cite[Lemma A.6]{OhnLin2023}.

Assumption \ref{ass:1} \textit{(i)}-\textit{(ii)} and \textit{(iv)}-\textit{(v)} are classical assumptions to get consistency of the posterior distribution. They guarantee that the prior charges the true value (wherever it is in the support) and any neighborhood of it. Assumption \ref{ass:1} \textit{(iii)} penalizes mixture models with a large number of components and further requires that the penalization becomes more severe as the sample size increases. A Gaussian prior distribution on $\beta$ satisfies Assumption \ref{ass:1} \textit{(iv)} under mild assumptions as we show in Lemma D.10 in the Supplementary Material.

To simplify notation, let $\mathcal{M}_{\leq K_N^{\st}} \equiv \mathcal{M}_{\leq K_N^{\st}}(\Theta)$. Our first theorem states that the posterior does not overestimate the number of components, that is, $\Pi(m\in\mathcal{M}_{\leq K^{\st}}|\by,\bZ,\by_0)$ converges to $1$ in $P_{N,0}^{\st}$ probability.
\begin{thm}\label{thm:1}
  Suppose that $\{y_{i,t}\}_t$ follows model \eqref{eq:2} with $|\gamma^{\st}| < 1$ and let the prior $\Pi$ satisfy Assumption \ref{ass:1} with $A>1$. Assume that $\theta_{j,\sigma^2}^{\st} \in [\ul{\sigma}^2, \ol{\sigma}^2]$ and $\theta_{j,\alpha}^{\st} \in [-L, L]$ for every $j\in[K^{\st}]$. 
  Then,
  \begin{equation}
    \Pi(K \leq K^{\st}|\by,\bZ,\by_0,N,v,\varphi,\varpi)\rightarrow 1
  \end{equation}
  in $P_{0}^{\st(N)}$-probability as $N\rightarrow \infty$.
\end{thm}
\indent In the next two sections, we establish convergence of the latent mixing measure with respect to the Wasserstein distance. We first consider the static panel data case and then the dynamic case. Here, we introduce some common notation. For some $K,K'\in\mathbb{N}$, consider a coupling $q$ of $\bw_K$ and $\bw_{K'}'$ defined as a joint distribution on $[1,\ldots,K] \times [1,\ldots,K']$ which is expressed as a $(K\times K')$-matrix $q = (q_{ij})_{1 \leq i\leq K,\, 1\leq j \leq K'}\in[0,1]^{K\times K'}$ and has marginal distributions $\sum_{i=1}^K q_{ij} = w_{j}'$ and $\sum_{j=1}^{K'} q_{ij} = w_{i}$ for every $i \in[K]$ and every $j\in [K']$. We denote by $Q(\bw_K,\bw_{K'}')$ the space of all such couplings. For every $\mathcal{q} \geq 1$, define the $\mathcal{q}$-th order Wasserstein distance between two atomic distributions $m:=\sum_{j=1}^K w_j \delta_{B_j,V_j}$ and $m':=\sum_{j=1}^{K'} w_j^{'} \delta_{B_j^{'},V_j'}$ with support in $\mathcal{B}\times \mathcal{V}$ as: for every $\mathcal{q} \geq 1$,
\begin{displaymath}
  W_{\mathcal{q}(m,m')} := \inf_{q\in Q(\bw_K,\bw_{K'}')}\left(\sum_{j=1}^K \sum_{h=1}^{K'} q_{jh}\rho^{\mathcal{q}}((B_j,V_j),(B_h',V_h'))\right)^{1/\mathcal{q}},
\end{displaymath}
\noindent where $\rho$ is a metric on $\mathcal{B}\times \mathcal{V}$.
The Wasserstein distance is less stringent than the Kolmogorov-Smirnov distance but at the same time is strong enough to provide meaningful guarantees on the means and weights. Wasserstein distance inherits the metric of the space of atomic support. So, if a mixing measure $m_N\rightarrow m$ with respect to the Wasserstein distance, then the ordered set of atoms of $m_N$ must converge to the atoms of $m$ in $\rho$ after permutation of atom labels.\\

\subsection{Posterior consistency in the static case}\label{ss:consistency:static}
Let us consider the static case where $h=0$ and the lagged dependent variable is not present in the model. In this case we use the notation $P_{m}^{(N)}$ and $P^{\st(N)}$ for $P_{m,0}^{(N)}$ and $P_{0}^{\st(N)}$, respectively. Suppose that $\bz_i$, $i\in [N]$, are \textit{i.i.d.} copies of $\bz$ which take values in $\mathbb{R}^{T\times p}$. We denote by $\iota_T$ the $T$-vector with all elements equal to one, $\Theta_{\alpha} := [-L,L]$ and $\Theta_{\sigma^2} := [\ul{\sigma}^2,\ol{\sigma}^2]$. We introduce the following class of functions:
\begin{multline*}
  \mathcal{B} := \Big\{(B_1(\cdot),\ldots,B_K(\cdot)):\mathbb{R}^{T\times p}\rightarrow \mathbb{R}^{T\times K}; \,\forall j\in[K],\, B_j(\bz) = \theta_{j,\alpha}\iota_T + \bz\beta, \\
  \theta_{j,\alpha} \in \Theta_{\alpha},\,\beta\in\mathbb{R}^p\Big\}.
\end{multline*}
\noindent Each element $(B_1,\ldots,B_K)$ in $\mathcal{B}$ is a $K$-vector of $T$-valued functions $B_j(\cdot)$ that associate $\bz\in\mathbb{R}^{T\times p}$ with a $T$-vector $\theta_{j,\alpha}\iota_T + \bz\beta$. The class $\mathcal{B}$ is indexed by $\btheta_{\alpha}$ and $\beta$. Let $\zeta:= \{\beta,\btheta_K,\bw_K,K\}$ be a $(p+2K+K+1)$-array of parameters taking values in $\mathcal{Z} := \mathbb{R}^p \times \Theta^{K}\times (0,1)^K\times\mathbb{N}_+$, where $\Theta = \Theta_{\alpha}\times \Theta_{\sigma^2}$. Let us consider the following finite multivariate conditional mixing distribution, conditional on $\bz$, with support points in $\mathcal{B}\times \Theta_{\sigma^2}$: for given $(B_1(\cdot),\ldots,B_K(\cdot)) \in \mathcal{B}$, $(\theta_{1,\sigma^2},\ldots,\theta_{K,\sigma^2})\in\Theta_{\sigma^2}^K$, $w_j >0$, for every $j\in[K]$, $\sum_{j=1}^K w_j = 1$, and $K\in\mathbb{N}_+$,
\begin{eqnarray*}
  \mathfrak{m}_{\bz} & \equiv & \mathfrak{m}_{\bz}(\mathbf{a},\sigma^2|\zeta,\cdot) := \sum_{j=1}^K w_j \delta_{B_j(\cdot),\theta_{j,\sigma^2}} (\mathbf{a},\sigma^2),\qquad \forall (\mathbf{a},\sigma^2)\in\mathbb{R}^{T}\times \mathbb{R}_+,
\end{eqnarray*}
\noindent where the subindex $\bz$ is used to stress the fact that this is a conditional distribution given $\bz$. Depending on the setting, the subindex can also denote the evaluation point of the conditioning variable: $\mfm_{\bz_i} \equiv \mfm_{\bz = \bz_i} \equiv \mathfrak{m}_{\bz}(\mathbf{a},\sigma^2|\zeta,\bz_i)$. 
This distribution has $K$ atoms and is an element of the set of multivariate conditional mixing measures, conditional on $\bz$, with exactly $K$ components:
\begin{multline*}
  \mathcal{M}_{K|\bz}(\wtl{\mathcal{Z}}) := \Big\{\sum_{j=1}^K w_j \delta_{B_j(\cdot),\theta_{j,\sigma^2}}(\cdot,\cdot), \, w_j >0, \sum_{j=1}^K w_j = 1, \theta_{j,\sigma^2} \in\Theta_{\sigma^2}, \forall j\in[K], \\
  (B_1(\cdot),\ldots,B_K(\cdot))\in\mathcal{B}\Big\},
\end{multline*}
\noindent where $\wtl{\mathcal{Z}} := \mathbb{R}^p \times \Theta^{K}\times (0,1)^K$ is the support of $\wtl\zeta_K := \{\beta,\btheta_K,\bw_K\}$. The conditioning on $\bz$ in $\mathcal{M}_{K|\bz}(\wtl{\mathcal{Z}})$ stresses the fact that the elements of this set are distributions conditional on $\bz$. 
The conditioning variables is the argument $\bz\in\mathbb{R}^{T\times p}$ of the functions $(B_1,\ldots, B_K)$.  There is a one-to-one correspondence between $\mathfrak{m}_{\bz}$ and $\zeta$ so that the prior on $\zeta$, specified in Section \ref{ss:prior}, defines the prior on $\mathfrak{m}_{\bz}$ conditional on $\bz$. \\
\indent By using the multivariate conditional mixing distribution $\mathfrak{m}_{\bz}$, the conditional joint distribution $P_{m}^{(N)}$ arising from the static version of model \eqref{eq:2} can be equivalently written as arising from the following multivariate model: for every $i=1,\ldots,N$,
  \begin{eqnarray}\label{eq:2:app:static}
    \by_{i} & = & \mathbf{a}_i + \mathbf{u}_{i},\qquad \mathbf{u}_{i}| \bz_{i} \sim \mathcal{N}_T(0,\Sigma_i),\qquad \Sigma_i = \Sigma(\sigma_i^2) \nonumber\\
    (\mathbf{a}_i,\sigma_i^2)|\bz_i & \sim & \mfm_{\bz}(\cdot,\cdot|\zeta,\bz_i).
  \end{eqnarray}
\noindent Hence, $P_{m}^{(N)} = P_{\mfm_{\bz}}^{(N)}:= \bigotimes_{i=1}^N P_{i,\mfm_{\bz}}$, where $P_{i,\mfm_{\bz}}$ denotes the conditional distribution of $\by_i$ given $(\zeta,\bz_i)$ according to model \eqref{eq:2:app:static}. Clearly, $P_{i,\mfm_{\bz}} = P_{i,m}$. 
The true model $P^{\st(N)} \equiv P_{\mfm_{\bz}^{\st}}^{(N)}$ is associated with the true multivariate conditional mixture measure $\mfm_{\bz}^{\st} \equiv \mfm_{\bz}(\alpha,\sigma^2|\zeta^{\st},\cdot)$.\\
\indent We denote by $\mathcal{M}_{\leq k|\bz}(\wtl{\mathcal{Z}}) := \bigcup_{j\leq k}\mathcal{M}_{j|\bz}(\wtl{\mathcal{Z}})$ the set of multivariate conditional mixing measures with at most $k$ components with finite support points in $\mathcal{B}\times \Theta_{\sigma^2}$ conditionally on $\bz$, and by $\mathcal{M}_{\bz}(\wtl{\mathcal{Z}}) := \bigcup_{k\in\mathbb{N}_+} \mathcal{M}_{k|\bz}(\wtl{\mathcal{Z}})$ the set of all multivariate conditional mixing distributions with finite support points in $\mathcal{B}\times \Theta_{\sigma^2}$ conditionally on $\bz$.\\
\indent Because $\mfm_{\bz}$ is a function of $\bz$, the Wasserstein distance between conditional distributions in $\mathcal{M}_{\bz}(\wtl{\mathcal{Z}})$ depends on $\bz$. We eliminate this dependence by considering the sample average, over the values of $\bz_i$ in the sample, of the Wasserstein norm of order $\mathcal{q}$ which we define as: for every $\mfm_{\bz}, \mfm_{\bz}'\in \mathcal{M}_{\bz}(\wtl{\mathcal{Z}})$,
$$\EE_N [W_{\mathcal{q}}(\mfm_{\bz},\mfm_{\bz}')] := \frac{1}{N}\sum_{i=1}^N W_{\mathcal{q}}(\mfm_{\bz_i},\mfm_{\bz_i}').$$
\noindent We consider the following Kullback-Leibler ball: $\forall \epsilon > 0$,
\begin{multline}
  B_{KL}^{\st}(\epsilon^2,\zeta^{\st},\mathcal{H};\bZ) := \Big\{\zeta\in\mathcal{Z}; \frac{1}{N}\sum_{i=1}^N KL(\zeta^{\st}, \zeta|\bz_{i}) \leq \epsilon^2 \log\left(\frac{1}{\epsilon}\right), \\
  \frac{1}{N}\sum_{i=1}^N KL_{2}(\zeta^{\st}, \zeta|\bz_{i}) \leq \epsilon^2 \left(\log\frac{1}{\epsilon}\right)^2\Big\}\label{def:KL:paper}
\end{multline}
with $KL(\zeta^{\st}, \zeta|\bz_{i}) := \mathcal{KL}(f_{\zeta^{\st}}(\cdot|\bz_{i})||f_{\zeta}(\cdot|\bz_{i}))$ and $KL_{2}(\zeta^{\st}, \zeta|\bz_i) := \mathcal{KL}_2(f_{\zeta^{\st}}(\cdot|\bz_{i})||f_{\zeta}(\cdot|\bz_{i}).$ and where $\mathcal{H}$ is defined in Online Appendix A.2.
We use the \emph{conditional Hellinger information of the $W_1$ metric for the subset} $\mathcal{M}_{\bz}(\wtl{\mathcal{Z}})$ which is defined as a real-valued function on the real line $\Psi_{\mathcal{M}_{\bz}(\wtl{\mathcal{Z}})}:\mathbb{R} \rightarrow \mathbb{R}$ as: for every $r>0$,
$$\Psi_{\mathcal{M}_{\bz}(\wtl{\mathcal{Z}})}(r) := \inf_{\mfm_{\bz} \in \mathcal{M}_{\bz}(\wtl{\mathcal{Z}}): \EE_N[W_1(\mfm_{\bz},\mfm_{\bz}^{\st})] \geq r/2} \EE_N\left[h^2(f_{\zeta}(\cdot|\bz),f_{\zeta^{\st}}(\cdot|\bz))\right].$$
The unconditional version of this notion has been introduced in \cite{Nguyen2013}. The function $r\mapsto \Psi_{\mathcal{M}_{\bz}(\wtl{\mathcal{Z}})}(r)$ is nonnegative and nondecreasing.\\
\indent The next theorem establishes posterior consistency for the mixing measure $\mfm_{\bz}$ with respect to the $W_1$-metric under three types of conditions. The first type involves the size of the support for the mixing measure (condition \eqref{condition:1}). It is quantified in terms of packing number. 
The second type of conditions is on the Hellinger information of the $W_1$ metric for the subset $\mathcal{M}_{\bz}(\wtl{\mathcal{Z}})$ which involves the likelihood of the model (conditions \eqref{app:condition:2} and \eqref{app:condition:3}). The third type of conditions is on the Kullback-Leibler support of the prior $\Pi$ and subsets of the space of discrete measures $\mathcal{M}_{\bz}(\wtl{\mathcal{Z}})$ (condition \eqref{app:condition:3} and \eqref{app:condition:4}). In Theorem \ref{thm:3:static} below we will use the explicit expression of the Hellinger information of the $W_1$ metric for the subset $\mathcal{M}_{\bz}(\wtl{\mathcal{Z}})$ and Assumption \ref{ass:1} as sufficient condition to guarantee conditions \eqref{app:condition:2} and \eqref{app:condition:3}. Recall the notation $D(\varepsilon,\mathcal{T},\rho)$ for the $\varepsilon$-packing number of the metric set $(\mathcal{T},\rho)$.

\begin{thm}\label{thm:2:static}
  Suppose that $\{y_{i,t}\}_t$ follows model \eqref{eq:2} without the lagged explanatory variable, $\theta_{j,\sigma^2}^{\st} \in [\ul{\sigma}^2, \ol{\sigma}^2]$ and $\theta_{j,\alpha}^{\st} \in [-L, L]$ for every $j\in[K^{\st}]$. Fix $\mfm_{\bz}^{\st} \in \mathcal{M}_{\bz}(\wtl{\mathcal{Z}})$, $\epsilon>0$, and consider a sequence of sets $\mathcal{G}_N \subseteq \mathcal{M}_{\bz}(\wtl{\mathcal{Z}})$ for which we define
  $$M(\mfm_{\bz},\Psi_{\mathcal{M}_{\bz}(\wtl{\mathcal{Z}})}(\epsilon)) := D\left(\frac{\Psi_{\mathcal{M}_{\bz}(\wtl{\mathcal{Z}})}^{1/2}(\epsilon)}{2},\mathcal{G}_N\cap \mathcal{U}(\mfm_{\bz},M_0\epsilon/2|\bZ), \sqrt{\EE_N[W_2^2(\cdot,\cdot)]}\right)$$
  for a given $\mfm_{\bz}\in \mathcal{G}_N$, for $\mathcal{U}(\mfm_{\bz},\epsilon|\bZ) := \{\wtl{\mfm}_{\bz}\in \mathcal{M}_{\bz}(\wtl{\mathcal{Z}}); \EE_N[W_1(\wtl{\mfm}_{\bz},\mfm_{\bz})] \leq \epsilon\}$, and for $M_0$ a positive constant. Let us assume that there are: non-negative sequences $\varepsilon_N \rightarrow 0$ and $C_N \equiv C_N(\bZ)>0$ such that either $N\varepsilon_N^2$ is bounded away from zero and $C_N\rightarrow \infty$ or $N\varepsilon_N^2\rightarrow \infty$ and $C_N$ is bounded, and such that the following holds: for every $\epsilon \geq \varepsilon_N$,
  \begin{multline}
    D\left(\frac{\epsilon}{2},\mathcal{G}_N \cap\left(\mathcal{U}(m_{\bz}^{\st},2C_N\epsilon|\bZ)\setminus \mathcal{U}(m_{\bz}^{\st},C_N\epsilon|\bZ)\right),\EE_N[W_1(\cdot,\cdot)]\right)\\
    \times \sup_{\mfm_{\bz}\in \mathcal{G}_N} M\left(\mfm_{\bz},\Psi_{\mathcal{M}_{\bz}(\wtl{\mathcal{Z}})}(\epsilon)\right)  \leq e^{N\varepsilon_N^2},\label{condition:1}
  \end{multline}
  \begin{align}
    & e^{C_N N\varepsilon_N^2 \log(1/\varepsilon_N)} \sum_{j\geq M_0} \exp\left\{-\frac{N}{48}\Psi_{\mathcal{M}_{\bz}(\wtl{\mathcal{Z}})}(j\varepsilon_N)\right\} \rightarrow 0;\label{app:condition:2}\\
    & \frac{\Pi(\mathcal{M}_{\bz}(\wtl{\mathcal{Z}})\cap \{m_{\bz}; \EE_N[W_1(\mfm_{\bz_i},\mfm_{\bz_i}^{\st})] \in [C_N j \varepsilon_N, 2C_N j \varepsilon_N]\})}{\Pi(B_{KL}^{\st}(\varepsilon_N^2,\zeta^{\st},\mathcal{H};\bZ))} \nonumber\\
    & \qquad \qquad \qquad \leq e^{N\Psi_{\mathcal{M}_{\bz}(\wtl{\mathcal{Z}})}(j\varepsilon_N)/48},\quad \forall j \geq M_0;\label{app:condition:3}\\ 
    & \frac{\Pi\left(\mathcal{M}_{\bz}(\wtl{\mathcal{Z}}) \setminus \mathcal{M}_{\leq K_N^{\st}|\bz}(\wtl{\mathcal{Z}}) \right)}{\Pi(B_{KL}^{\st}(\varepsilon_N^2,\zeta^{\st},\mathcal{H};\bZ))} = o\left(e^{-C_N N\varepsilon_N^2\log(1/\varepsilon_N)}\right).\label{app:condition:4}
  \end{align}

  Then,
  \begin{equation}
    \Pi\left(\left.\mfm_{\bz}\in \mathcal{M}_{\bz}(\wtl{\mathcal{Z}}); \EE_N[W_1(\mfm_{\bz},\mfm_{\bz}^{\st})] \geq C_N M_0 \varepsilon_N\right| \by,\bZ,N,v,\varphi,\varpi\right) \rightarrow 0
  \end{equation}
  in $P^{\st(N)}$-probability.
\end{thm}

The next theorem establishes posterior consistency under Assumption \ref{ass:1} \textit{(i)}-\textit{(iv)} under which we can prove that conditions \eqref{app:condition:3}-\eqref{app:condition:4} of Theorem \ref{thm:2:static} hold. Condition \eqref{app:condition:2} can be directly checked by using the explicit expression of the Hellinger information of the $W_1$ metric for the subset $\mathcal{M}_{\bz}(\wtl{\mathcal{Z}})$.

\begin{thm}\label{thm:3:static}
  Suppose that $\{y_{i,t}\}_t$ follows model \eqref{eq:2} without the lagged dependent variable and let the prior $\Pi$ satisfy Assumption \ref{ass:1} \textit{(i)}-\textit{(iv)} with $A>1$. Assume that \textit{(i)} $\theta_{j,\sigma^2}^{\st} \in [\ul{\sigma}^2, \ol{\sigma}^2]$, \textit{(ii)} $\theta_{j,\alpha}^{\st} \in [-L, L]$ for every $j\in[K^{\st}]$, and \textit{(iii)} $\Pi(K = k|N)\gtrsim N^{-c}$ for every $k \in \mathbb{N}$ and for some constant $c > 0$. Moreover, assume that condition \eqref{condition:1} in Theorem \ref{thm:2:static} holds. Then, for every sequence $C_N \rightarrow\infty$ 
  \begin{equation}
    \Pi\left(\left.\mfm_{\bz} \in \mathcal{M}_{\bz}(\wtl{\mathcal{Z}}); \EE_N[W_1(\mfm_{\bz_i},\mfm_{\bz_i}^{\st})] \geq C_N\sqrt{\log(N)/N}\right| \by,\bZ,N,v,\varphi,\varpi\right) \rightarrow 0
  \end{equation}
  in $P^{\st(N)}$-probability.
\end{thm}
%

\subsection{Posterior consistency in the dynamic case}\label{ss:consistency:dynamic}
In this section we consider the dynamic case \eqref{eq:2} where the lagged value of the dependent variable is among the covariates. Suppose that $\bz_i$, $i \in [N]$, are \textit{i.i.d.} copies of $\bz$ which takes values in $\mathbb{R}^{T\times p}$. We denote by $\iota_T$ the $T$-vector with all elements equal to one, $\Theta_{\alpha} := [-L,L]$, $\Theta_{\sigma^2} := [\ul{\sigma}^2,\ol{\sigma}^2]$, and $\gamma^{[1:T]}:=(\gamma,\gamma^2,\gamma^3,\ldots,\gamma^T)'$. Moreover, $\Gamma$ denotes a $(T\times T)$-lower triangular Topelitz matrix with one on its main diagonal, that is, $\Gamma =(\Gamma_{i,j})_{i,j}$ and $\Gamma_{i,j} = \gamma^{|i-j|}$ if $i\geq j$ and $\Gamma_{i,j}=0$ otherwise. Therefore, the $T$-vector $\Gamma \bz_i \beta$ has $t$-th element $\beta' \sum_{\ell = 0}^{t-1}\gamma^{\ell}z_{i,t-\ell-h}$ for $t=1,\ldots,T$. Similarly as in Section \ref{ss:consistency:static} we introduce the following class of functions:
\begin{multline*}
  \mathcal{B}_d := \Big\{(B_1(\cdot),\ldots,B_K(\cdot)):\mathbb{R}^{T\times p}\times (-1,1)\rightarrow \mathbb{R}^{T\times K}; \,\forall j\in[K],\\
  B_j(\bz,y_0) = \frac{\theta_{j,\alpha}}{1 - \gamma}(\iota_T - \gamma^{[1:T]}) + \gamma^{[1:T]} y_{0} + \Gamma \bz_i \beta, \;
  \theta_{j,\alpha} \in \Theta_{\alpha},\,\beta\in\mathbb{R}^p, \gamma\in(-1,1)\Big\}.
\end{multline*}
\noindent Each element $(B_1,\ldots,B_K)$ in $\mathcal{B}_d$ is a $K$-vector of $T$-valued functions $B_j(\cdot)$ that associate $\bz\in\mathbb{R}^{T\times p}$ and $y_{0}$ with a $T$-vector $\frac{\theta_{j,\alpha}}{1 - \gamma}(\iota_T - \gamma^{[1:T]}) + \gamma^{[1:T]} y_{0} + \Gamma \bz_i \beta$. The class $\mathcal{B}_d$ is indexed by $\btheta_{\alpha}$, $\gamma$ and $\beta$. Let $\zeta:= \{\gamma,\beta,\btheta_K,\bw_K,K\}$ be a $(3K+p+2)$-array of parameters taking values in $\mathcal{Z}_d := (-1,1) \times \mathbb{R}^p \times \Theta^{K}\times (0,1)^K \times \mathbb{N}_+$, where $\Theta = \Theta_{\alpha}\times \Theta_{\sigma^2}$. Let us consider the following finite multivariate conditional mixing distribution with support points in $\mathcal{B}_d \times \Theta_{\sigma^2}$ conditional on $(\bz,y_0)$: for given $(B_1(\cdot),\ldots,B_K(\cdot)) \in \mathcal{B}_d$, $(\theta_{1,\sigma^2},\ldots,\theta_{K,\sigma^2})\in\Theta_{\sigma^2}^K$, $w_j>0$ for every $j\in[K]$, $\sum_{j=1}^K w_j = 1$, and $K\in\mathbb{N}_+$,
\begin{eqnarray*}
  \mathfrak{m}_{\bz 0} & \equiv & \mathfrak{m}_{\bz 0}(\mathbf{a},\sigma^2|\zeta,\cdot) := \sum_{j=1}^K w_j \delta_{B_j(\cdot),\theta_{j,\sigma^2}} (\mathbf{a},\sigma^2),\qquad \forall (\mathbf{a},\sigma^2)\in\mathbb{R}^{T}\times \mathbb{R}_+,
\end{eqnarray*}
\noindent where the subindex $\bz 0$ is used to stress the fact that this is a conditional distribution given $(\bz,y_0)$. Depending on the setting, the subindex can also denote the evaluation point of the conditioning variable: $\mfm_{\bz_i 0} \equiv \mfm_{\bz = \bz_i, y_0 = y_{i0}} \equiv \mathfrak{m}_{\bz 0}(\mathbf{a},\sigma^2|\zeta,\bz_i, y_{i0})$. This distribution has $K$ atoms and is an element of the set of multivariate conditional mixing measures, conditional on $(\bz, y_0)$, with exactly $K$ components:
\begin{multline*}
  \mathcal{M}_{K|\bz 0}(\wtl{\mathcal{Z}}_d) := \Big\{\sum_{j=1}^K w_j \delta_{B_j(\cdot),\theta_{j,\sigma^2}} (\cdot, \cdot), \, w_j >0, \sum_{j=1}^k w_j = 1, \theta_{j,\sigma^2} \in\Theta_{\sigma^2}, \forall j\in[K], \\
  (B_1(\cdot),\ldots,B_K(\cdot))\in\mathcal{B}_d\Big\},
\end{multline*}
\noindent where $\wtl{\mathcal{Z}}_d := (-1,1) \times \mathbb{R}^p \times \Theta^{K}\times [0,1]^K$ is the support of $\wtl\zeta_K := \{\gamma,\beta,\btheta_K,\bw_K\}$. The conditioning on $\bz 0$ in $\mathcal{M}_{K|\bz 0}(\wtl{\mathcal{Z}}_d)$ stresses the fact that the elements of this set are distributions conditional on $(\bz, y_0)$. There is a one-to-one correspondence between $\mathfrak{m}_{\bz 0}$ and $\zeta$ so that the prior on $\zeta$ defines the prior on $\mathfrak{m}_{\bz 0}$ conditional on $(\bz, y_0)$.\\
\indent As for the static case, the conditional joint distribution $P_{m,0}^{(N)}$ arising from the dynamic model \eqref{eq:2} can be equivalently written as arising from the following multivariate model: for every $i=1,\ldots,N$,
  \begin{eqnarray}
    \by_{i} & = & \mathbf{a}_i + \mathbf{u}_{i},\qquad \mathbf{u}_{i}| \bz_{i} \sim \mathcal{N}_T(0,\Sigma_i),\qquad \Sigma_i = \Sigma(\sigma_i^2)  \nonumber\\
    (\mathbf{a}_i,\sigma_i^2)|\bz_i & \sim & \mfm_{\bz 0}(\cdot,\cdot|\zeta,\bz_i,y_{i0}). \label{eq:2:app}
  \end{eqnarray}
\noindent Hence, $P_{m,0}^{(N)} = P_{\mfm_{\bz 0}}^{(N)}:= \bigotimes_{i=1}^N P_{i,\mfm_{\bz 0}}$, where $P_{i,\mfm_{\bz 0}} $ denotes the conditional distribution of $\by_i$ given $(\zeta,\bz_i, y_{i0})$ according to model \eqref{eq:2:app}. Clearly, $P_{i,\mfm_{\bz 0}} = P_{i, m, 0}$. The true conditional model $P_{0}^{\st(N)} \equiv P_{\mfm_{\bz 0}^{\st(N)}}$ is associated with the true multivariate conditional mixture measure $\mfm_{\bz 0}^{\st} \equiv \mfm_{\bz 0} (\alpha,\sigma^2|\zeta^{\st},\cdot, \cdot)$.\\
\indent Similarly as in Section \ref{ss:consistency:static}, we denote by $\mathcal{M}_{\leq k|\bz 0}(\wtl{\mathcal{Z}}_d) := \bigcup_{j\leq k}\mathcal{M}_{j|\bz 0}(\wtl{\mathcal{Z}}_d)$ the set of multivariate conditional mixing measures with at most $k$ components, and by $\mathcal{M}_{\bz 0}(\wtl{\mathcal{Z}}_d) := \bigcup_{k\in\mathbb{N}_+} \mathcal{M}_{k|\bz 0}(\wtl{\mathcal{Z}}_d)$ the set of multivariate conditional mixing distributions with finite support points in $\mathcal{B}_d\times \Theta_{\sigma^2}$ conditionally on $(\bz,y_{0})$.\\
%
\indent A theorem equivalent to Theorem \eqref{thm:2:static} holds for the dynamic model. We postpone it to Online Appendix A.2.5 to shorten the manuscript. Instead, we present here the result of posterior consistency with respect to the average Wasserstein norm of order $1$. The average Wasserstein norm of order $q$ is defined as: for every $\mfm_{\bz 0}, \mfm_{\bz 0}'\in \mathcal{M}_{\bz 0}(\wtl{\mathcal{Z}}_d)$
$$\EE_N [W_{\mathcal{q}}(\mfm_{\bz 0},\mfm_{\bz 0}')] := \frac{1}{N}\sum_{i=1}^N W_{\mathcal{q}}(\mathfrak{m}_{\bz_i 0},\mathfrak{m}_{\bz_i 0}').$$
Recall the notation $D(\varepsilon,\mathcal{T},\rho)$ for the $\varepsilon$-packing number of the metric set $(\mathcal{T},\rho)$.
\begin{thm}\label{thm:3:dynamic}
  Suppose that $\{y_{i,t}\}_t$ follows model \eqref{eq:2} with $|\gamma^{\st}| < 1$ and let the prior $\Pi$ satisfy Assumption \ref{ass:1} with $A>1$. Assume that \textit{(i)} $\theta_{j,\sigma^2}^{\st} \in [\ul{\sigma}^2,  \ol{\sigma}^2]$, \textit{(ii)} $\theta_{j,\alpha}^{\st} \in [-L, L]$ for every $j\in[K^{\st}]$, and \textit{(iii)} $\Pi(K = k|N)\gtrsim N^{-c}$ for every $k \in \mathbb{N}$ and for some constant $c > 0$. Fix $\mfm_{\bz 0}^{\st} \in \mathcal{M}_{\bz 0}(\wtl{\mathcal{Z}}_d)$, $r>0$, and consider a sequence of sets $\mathcal{G}_N \subseteq \mathcal{M}_{\bz 0}(\wtl{\mathcal{Z}}_d)$ for which we define
  $$M(\mfm_{\bz 0},\Psi_{\mathcal{M}_{\bz 0}(\wtl{\mathcal{Z}}_d)}(r)) := D\left(\frac{\Psi_{\mathcal{M}_{\bz 0}(\wtl{\mathcal{Z}}_d)}^{1/2}(r)}{2},\mathcal{U}(\mfm_{\bz 0},M_0 r /2|\bZ, \by_0), \sqrt{\EE_N[W_2^2(\cdot,\cdot)]}\right)$$
  for a given $\mfm_{\bz 0}\in \mathcal{G}_N$, $M_0$ a positive constant, and where $\mathcal{U}(\mfm_{\bz 0},r|\bZ, \by_0) := \{\wtl{\mfm}_{\bz 0}\in \mathcal{G}_N; \EE_N[W_1(\wtl{\mfm}_{\bz 0},\mfm_{\bz 0})] \leq r\}$. Let us assume that there is a non-negative sequences $C_N\rightarrow \infty$ such that: for every $\epsilon \geq N^{-1/2}$,
  \begin{multline}\label{condition:1:dynamic}
    D\left(\frac{\epsilon}{2},\mathcal{G}_N(\mathcal{U}(m_{\bz 0}^{\st},2C_N\epsilon|\bZ,\by_0)\setminus \mathcal{U}(m_{\bz 0}^{\st},C_N \epsilon|\bZ, \by_0)),\EE_N[W_1(\cdot,\cdot)]\right)\\
    \times \sup_{\mfm_{\bz 0}\in \mathcal{G}_N} M(\mfm_{\bz 0},\Psi_{\mathcal{M}_{\bz 0}(\wtl{\mathcal{Z}}_d)}(\epsilon)) \lesssim e.
  \end{multline}
  Then,
  \begin{equation}
    \Pi\left(\left.\mfm_{\bz 0}\in \mathcal{M}_{\bz 0}(\wtl{\mathcal{Z}}_d); \EE_N[W_1(\mfm_{\bz_i 0},\mfm_{\bz_i 0}^{\st})] \geq C_N\sqrt{\log(N)/N}\right| \by,\bZ,\by_0,N,v,\varphi,\varpi\right) \rightarrow 0
  \end{equation}
  in $P^{\st(N)}$-probability.
\end{thm}

\section{Numerical experiment}\label{s:Monte:Carlo}
In this section we study finite sample properties of our Bayesian procedure by using simulated data and the \emph{telescoping sampling} described in Algorithm \ref{algorithm:1}. The details of the implementation are presented in Section \ref{ss:sim:implementation}. In Sections \ref{ss:results}-\ref{ss:sim:covariates} we present the results of the Monte Carlo exercise. We consider two setting: the static case where no lagged dependent variable is present among the explanatory variables, and the dynamic case where the lagged dependent variable is included. Then, we consider the impact of not including relevant covariates on the ability of detecting the clustering structure.

\subsection{Implementation}\label{ss:sim:implementation}
Data are generated by using model \ref{eq:distribution:chi}-\ref{eq:distribution:conditional:chi}. In the static case, the lagged dependent variable is not in the model and $\beta^{\st}$ is set equal to zero. In the dynamic case, we set $\beta^{\st} = 0$ and $\gamma^{\st} = 0.1$. According with Section \ref{ss:prior}, we specify the prior as:
\begin{align*}
  (\gamma,\beta)|\varpi & \sim \mathcal{N}(\gamma_0,\Gamma_0;-1,1)\mathcal{N}_p(\beta_0,\Omega_0),\qquad &\textrm{ with }\varpi_1 = \{\gamma_0,\Gamma_0\} \textrm{ and }\varpi_2 = \{\beta_0,\Omega_0\},\\
  \theta_k|\varphi & \sim \mathcal{N}(b_0,B_0)\mathcal{IG}(c_0,C_0),  &\textrm{ independently for } k \in[K], \\
  & &\textrm{ with }\varphi_1 = \{b_0,B_0\}\textrm{ and }\varphi_2 = \{c_0,C_0\},\\
  C_0 & \sim \mathcal{G}(g_0,G_0),&\\
  \bw_K|K,v & \sim \mathcal{D}ir(v,\ldots,v),\qquad &\textrm{either } v = e_0,  \textrm{ or } v=\frac{e_0}{K},\\
  &&\textrm{either } e_0 = 1, \textrm{ or } e_0 \sim\mathcal{G}(1,20),\\
  K-1 & \sim BNB(a_{\lambda},a_{\pi},b_{\pi}), &
\end{align*}
\noindent where $\mathcal{N}(\gamma_0,\Gamma_0;-1,1)$ denotes a truncated Normal distribution with mean $\gamma_0$, variance $\Gamma_0$, truncated on $(-1,1)$, $\mathcal{G}(\cdot,\cdot)$ denotes the Gamma distribution,  $\mathcal{IG}(\cdot, \cdot)$ the inverse gamma distribution, $\mathcal{D}ir(v,\ldots,v)$ denotes the symmetric Dirichlet distribution with concentration parameter $v >0$, and $BNB(\cdot,\cdot,\cdot)$ denotes a beta-negative-binomial distribution (see Supplementary Material F.1). Because $\bw_K|K,v\sim\mathcal{D}ir(v,\ldots,v)$, then the prior mean of an element $w_k$ of $\bw_K$ is $K^{-1}$ for every $k\in[K]$. The prior variance of $w_k$ is $Var(w_k|K,v) = \frac{K - 1}{K^2(v K + 1)}$ for every $k\in[K]$ and it decreases with $v$. This means that a large value of $v$ favour vectors $\bw_K$ with balanced components. As discussed in Section \ref{ss:prior}, if $v$ is equal to a value $e_0$ we have a static MFM, if $v=e_0/K$ we have a dynamic MFM. For both the static and dynamic MFM, in our simulation we have tried the parameter $e_0$ fixed to $1$ -- in which case the symmetric Dirichlet distribution is equivalent to a uniform distribution over all points in its support (flat Dirichlet distribution), and the hyperparameter $e_0$ drawn from a Gamma distribution: $e_0\sim\mathcal{G}(1,20)$, where $1$ is the shape parameter and $20$ the rate parameter.\\
\indent The prior on $K$ is constructed starting from the translated Poisson distribution $K-1\sim\mathcal{P}oi(\lambda)$ introduced by \cite{MillerHarrison2018} where $\lambda$ is integrated out based on the gamma distribution $\lambda\sim \mathcal{G}(a_{\lambda},\pi)$. The resulting prior is negative binomial: $K-1 \sim NegBin(a_{\lambda},\pi)$, and then we integrate out $\pi$ with respect to a Beta distribution $\pi\sim\mathcal{B}eta(a_{\pi},b_{\pi})$. This integration yields that marginally $K-1$ has a beta-negative-binomial (BNB) distribution: $K-1 \sim BNB(a_{\lambda},a_{\pi},b_{\pi})$ (see Supplementary Material F.1 for more details). We have tried different values of these parameters in our implementation. In addition we have tried as alternative priors: a geometric prior distribution (with success probabilities $0.5$, $0.2$ or $0.1$), a uniform distribution over a fixed interval, a Poisson distribution with rate $1$, $4$ or $9$, a Negative Binomial prior distribution with probability $0.5$ and size $1$, $4$, or $9$, a degenerate distribution on a fixed $\ol{K}$. The results are quite robust to these different specifications.

In each draw of our MCMC, to identify the atoms of the cluster we use two alternative post-processing strategies. Both determine a unique labeling of the MCMC draws after selecting a number of cluster, which is chosen in our case based on the mode of the posterior of $K_+$. The first identification strategy is based on the ordering constraints: $\theta_{1,\alpha} < \theta_{2,\alpha} < \ldots < \theta_{K,\alpha},$
\noindent which solve the identification issue due to label switching, see \textit{e.g.} \cite{FruwirthSchnatter2006}. The other components of $\btheta_K$, the weights and the latent allocation variables are then reordered accordingly. The second identification strategy that we use is based on clustering the $\theta_{k,\alpha}$ in the point processing representation (\cite{FruwirthSchnatter2006}). We describe this strategy in Supplementary Material F.3.

\subsection{Results of the Monte Carlo simulation}\label{ss:results}
We have run $100$ Monte Carlo (MC) iterations and for each of these iterations we have run the telescoping sampler algorithm \ref{algorithm:1} with $10,000$ MCMC iterations after $100$ iterations of burn-in period. We have tried different number of clusters in the population: $K^{\st}\in\{3,9\}$, different values of $\bw_K$ and $\btheta_K$, and different values for $N$ and $T$: $N\in\{50,100,500\}$, $T\in\{3,30\}$.\\
\indent For each Monte Carlo iteration, we estimate $K$ and $K_+$ as the maximum a-posteriori (that is, the most frequent value among the $10,000$ MCMC draws from the posteriors of $K$ and $K_+$), denoted as $\wh{K}^{(m)}$ and $\wh{K}_+^{(m)}$ for the $m$-th iteration. Then, we take the average of these values across the $100$ Monte Carlo iterations, and we denote them as $\wh{K}$ and $\wh{K}_+$. We also compute the first and third quartiles of the posterior of $K$ and $K_+$ for each MC iteration and then take the average over the $100$ MC iterations. To estimate the atoms and their weights, for each Monte Carlo iteration we compute their posterior means. Then, if there is at least one Monte Carlo iteration with a number of clusters equal to $\wh{K}_+$, we take the average over only the Monte Carlo iterations with a number of clusters less than or equal to $\wh{K}_+$. On the other hand, if there is no Monte Carlo iteration with a number of clusters equal to $\wh{K}_+$, then we take the average over only the Monte Carlo iterations with a number of clusters equal to the true value $K^*$. This second case is rare and we have experienced it only when $K^*$ is large, $K^* = 9$, and the model is dynamic.\\
\indent Let us start with considering the static case where there is no lagged dependent variable $y_{i,t-1}$ and where the true value of $\beta$ has been set equal to $0$. The results are reported in Tables \ref{tab:1:static:MC}-\ref{tab:3:static:MC}. In Table \ref{tab:1:static:MC} we study the effect of augmenting $N$ and $T$ in the case where the atoms $\theta_{1,\alpha},\ldots,\theta_{K,\alpha}$ are well separated while the other atoms $\theta_{1,\sigma^2},\ldots, \theta_{K,\sigma^2}$ have the same value $1$. We fix $K^{\st} = 3$ and all the three components have the same weights. The results show that the atoms, the weight, $K^{\st}$ and $\beta^{\st}$ are very well estimated even for small values of $N$ and $T$. The effect of increasing $N$ and $T$ is negligible in this case. On the other hand, when two elements of the atoms $\theta_{1,\alpha},\ldots,\theta_{K,\alpha}$ are very closed (that is, $0$ and $0.5$ in our simulation), then Table \ref{tab:2:static:MC} shows that we cannot recover the true $K^{\st}$ even with a relatively large $N$ if $T$ is small. A slightly larger $T$ ($T=100$) instead, allows to perfectly recover the group structure and the atoms. This shows the usefulness of panel data in order to recover the mixture structure. Estimation of $\beta^{\st}$ is always very good, even for small $N$.\\
\indent Finally, Table \ref{tab:3:static:MC} shows the results of our procedure when the number of components is large, that is, $K^{\st} = 9$. In this case, our procedure slightly overestimates $K^{\st}$ when $N$ is small by providing the estimates $\wh{K} = 12$ and $\wh{K}_+ = 10$. By increasing $N$ from $50$ to $100$ the results improve and obtain an estimate equal to the true $K^{\st}$: $\wh{K}_+ = 9$. The atoms and the weights of the components are perfectly estimated. We have tried different values of $T$ and they have no impact.\\
\indent The general message is that in static models with finite samples, when at least one component of the atoms varies sufficiently across the groups, then we can recover the group structure very well even with a very small number of periods ($T=3$). As long as the variation in the atoms is minimal, then we need a $T$ large to recover the group structure with a finite $N$. This is the benefit of considering panel data.

{\footnotesize
\begin{table}[!h]
\centering
\begin{tabular}{|c|c|c|c||c|c|c|c||}
  \hline
  N & T & $\theta_{\alpha}^{\star},\theta_{\sigma^2}^{\star}$ & $w^{\star}$  & $\wh\theta$ & $\wh w_K$  & $\wh\beta$ & $\widehat{K} $, $\widehat{K}_+ $\\
  \hline
  \hline
  $50$ & $3$ & $\begin{array}{cc}-5 & 1\\ 0 & 1\\ 5 & 1\end{array}$ & $\begin{array}
    {c} 1/3 \\ 1/3 \\ 1/3 \end{array}$ & $\begin{array}{cc}-5.03 & 1.06\\ 0.02 & 1.08\\ 5.00 & 1.03\end{array}$ & $\begin{array}
    {c} 0.33 \\ 0.33 \\ 0.34
  \end{array}$ & $\begin{array}
    {c} -0.05\\ \scriptstyle{(-0.28,0.18)}\\
  \end{array}$ & $\begin{array}
    {c l}
    \wh K: & \; 3\\ & \scriptstyle{(3,4)}\\
    \wh K_+: & \; 3\\ & \scriptstyle{(3,3)}
  \end{array}$\\
  \hline
  $50$ & $30$ & $\begin{array}{cc}-5 & 1\\ 0 & 1\\ 5 & 1\end{array}$ & $\begin{array}
    {c} 1/3 \\ 1/3 \\ 1/3 \end{array}$ & $\begin{array}{cc}-5.00 & 1.00\\ 0.01 & 1.00\\ 5.00 & 1.00\end{array}$ & $\begin{array}
    {c} 0.34 \\ 0.34 \\ 0.32
  \end{array}$ & $\begin{array}
    {c} -0.03\\ \scriptstyle{(-0.09,0.02)}\\
  \end{array}$ & $\begin{array}
    {c l}
    \wh K: & \; 3\\ & \scriptstyle{(3,3)}\\
    \wh K_+: & \; 3\\ & \scriptstyle{(3,3)}
  \end{array}$\\
  \hline
  $100$ & $3$ & $\begin{array}{cc}-5 & 1\\ 0 & 1\\ 5 & 1\end{array}$ & $\begin{array}
    {c} 1/3 \\ 1/3 \\ 1/3 \end{array}$ & $\begin{array}{cc}-4.99 & 1.03\\ 0.01 & 1.02\\ 5.01 & 1\end{array}$ & $\begin{array}
    {c} 0.33 \\ 0.33 \\ 0.33
  \end{array}$ & $\begin{array}
    {c} 0.03\\ \scriptstyle{(-0.11,0.17)}\\
  \end{array}$ & $\begin{array}
    {c l}
    \wh K: & \; 3\\ & \scriptstyle{(3,3)}\\
    \wh K_+: & \; 3\\ & \scriptstyle{(3,3)}
  \end{array}$\\
  \hline
  $500$ & $3$ & $\begin{array}{cc}-5 & 1\\ 0 & 1\\ 5 & 1\end{array}$ & $\begin{array}
    {c} 1/3 \\ 1/3 \\ 1/3 \end{array}$ & $\begin{array}{cc}-4.99 & 1.00\\ 0.00 & 1.00\\ 5.00 & 0.99\end{array}$ & $\begin{array}
    {c} 0.34 \\ 0.34 \\ 0.33
  \end{array}$ & $\begin{array}
    {c} -0.03\\ \scriptstyle{(-0.09,0.02)}\\
  \end{array}$ & $\begin{array}
    {c l}
    \wh K: & \;3\\ & \scriptstyle{(3,3)}\\
    \wh K_+: & \;3\\ & \scriptstyle{(3,3)}
  \end{array}$\\
  \hline
\end{tabular}
\caption{{\footnotesize Static case. Results of a Monte Carlo exercise with $100$ iterations. Study of the impact of increasing $T$ and/or $N$. $\beta^{\star} = 0$, $K^{\star} = 3$. The estimation are means across the $100$ Monte Carlo iterations of the posterior means. The credible intervals (CI) for $\beta$ are the $95\%$ CI, and the $1^{st}$ and $3^{rd}$ quartiles for $\wh K$ and $\wh K_+$.}}\label{tab:1:static:MC}
\end{table}
}
\normalsize
{\footnotesize
\begin{table}[!h]
\centering
\begin{tabular}{|c|c|c|c||c|c|c|c||}
  \hline
  N & T & $\theta_{\alpha}^{\star},\theta_{\sigma^2}^{\star}$ & $w^{\star}$  & $\wh\theta$ & $\wh w_K$  & $\wh\beta$ & $\widehat{K} $, $\widehat{K}_+ $\\
  \hline
  \hline
  $50$ & $3$ & $\begin{array}{cc}-5 & 1\\ 0 & 1\\ 0.5 & 1\end{array}$ & $\begin{array}
    {c} 0.45 \\ 0.5 \\ 0.05 \end{array}$ & $\begin{array}{cc}-5.01 & 1.07\\ 0.05 & 1.05\end{array}$ & $\begin{array}
    {c} 0.45 \\0.55
  \end{array}$ & $\begin{array}
    {c} \wh\beta: -0.03\\ \scriptstyle{(-0.22,0.15)}
  \end{array}$ & $\begin{array}
    {cl}
    \wh K: & \; 2\\ & \scriptstyle{(2,2)}\\
    \wh K_+: & \; 2\\ & \scriptstyle{(2,2)}
  \end{array}$\\
  \hline
  $50$ & $3$ & $\begin{array}{cc}-5 & 1\\ 0 & 1\\ 0.5 & 1\end{array}$ & $\begin{array}
    {c} 1/3 \\ 1/3 \\ 1/3 \end{array}$ & $\begin{array}{cc}-5.03 & 1.07\\ 0.26 & 1.09\end{array}$ & $\begin{array}
    {c} 0.33 \\0.67
  \end{array}$ & $\begin{array}
    {c} \wh\beta: -0.02\\ \scriptstyle{(-0.21,0.17)}
  \end{array}$ & $\begin{array}
    {cl}
    \wh K: & \; 2\\ & \scriptstyle{(2,2)}\\
    \wh K_+: & \; 2\\ & \scriptstyle{(2,2)}
  \end{array}$\\
  \hline
  $500$ & $3$ & $\begin{array}{cc}-5 & 1\\ 0 & 1\\ 0.5 & 1\end{array}$ & $\begin{array}
    {c} 0.45 \\ 0.5 \\ 0.05 \end{array}$ & $\begin{array}{cc}-4.99 & 1\\ 0.05 & 1.01 \end{array}$ & $\begin{array}
    {c} 0.45 \\0.55
  \end{array}$ & $\begin{array}
    {c} \wh\beta: -0.01\\ \scriptstyle{(-0.07,0.04)}
  \end{array}$ & $\begin{array}
    {cl}
    \wh K: & \; 2\\ & \scriptstyle{(2,2)}\\
    \wh K_+: & \; 2\\ & \scriptstyle{(2,2)}
  \end{array}$\\
  \hline
  $50$ & $100$ & $\begin{array}{cc}-5 & 1\\ 0 & 1\\ 0.5 & 1\end{array}$ & $\begin{array}
    {c} 0.45 \\ 0.5 \\ 0.05 \end{array}$ & $\begin{array}{cc}-4.99 & 1\\ -0.00 & 1\\ 0.51 & 1 \end{array}$ & $\begin{array}
    {c} 0.45 \\0.47 \\ 0.07
  \end{array}$ & $\begin{array}
    {c} \wh\beta: -0.00\\ \scriptstyle{(-0.03,0.02)}
  \end{array}$ & $\begin{array}
    {cl}
    \wh K: & \; 3\\ & \scriptstyle{(3,3)}\\
    \wh K_+: & \; 3\\ & \scriptstyle{(3,3)}
  \end{array}$\\
  \hline
\end{tabular}
\caption{{\footnotesize Static case. Results of a Monte Carlo exercise with $100$ iterations. Study of the impact when we have two components of $\theta_{\alpha}$ and of $\theta_{\sigma^2}$ very closed and/or different weights across components. $\beta^{\star} = 0$, $K^{\star} = 3$. The estimation are means across the $100$ Monte Carlo iterations of the posterior means. The credible intervals (CI) for $\beta$ are the $95\%$ CI, and the $1^{st}$ and $3^{rd}$ quartiles for $\wh K$ and $\wh K_+$.}}\label{tab:2:static:MC}
\end{table}
}
\normalsize

{\footnotesize
\begin{table}[!h]
\centering
\begin{tabular}{|l|c|c||c|c|c||}
  \hline
  N & $\theta_{\alpha}^{\star},\theta_{\sigma^2}^{\star}$ & $w^{\star}$  & $\wh\theta$ & $\wh w_K$  & $\wh\beta$, $\widehat{K} $, $\widehat{K}_+ $\\
  \hline
  \hline
  $50$ & $\begin{array}{cc}-20 & 1\\ -15 & 1\\ -10 & 1\\ -5 & 1\\ 0 & 1\\ 5 & 2 \\ 10& 0.5\\ 15 & 0.5\\ 20 & 0.5\end{array}$ & $\begin{array}
    {c} 0.11 \\ 0.11 \\ 0.11 \\ 0.11 \\ 0.11 \\ 0.11 \\ 0.11 \\ 0.11 \\ 0.11
  \end{array}$ & $\begin{array}{cc}-20 & 1.06\\ -15.13 & 1.15\\ -10.28 & 1.15\\ -5.55 & 1.21\\ -0.90 & 1.02\\ 3.76 & 1.30 \\ 8.58 & 1.87\\ 13.40& 0.62\\ 18.22 & 0.65\\ 20.02 & 0.62
  \end{array}$ & $\begin{array}
    {c} 0.09 \\ 0.09 \\ 0.10 \\ 0.09 \\ 0.10 \\ 0.10 \\ 0.10 \\ 0.10 \\ 0.11 \\ 0.12
  \end{array}$ & $\begin{array}
    {c} \wh\beta: -0.14\\ \scriptstyle{(-1.13,0.84)}\\
    \\
    \wh K: 12\\ \scriptstyle{(10,13)}\\
    \\
    \wh K_+: 10\\ \scriptstyle{(9,11)}
  \end{array}$\\
  \hline
    $100$ & $\begin{array}{cc}-20 & 1\\ -15 & 1\\ -10 & 1\\ -5 & 1\\ 0 & 1\\ 5 & 2 \\ 10& 0.5\\ 15 & 0.5\\ 20 & 0.5\end{array}$ & $\begin{array}
    {c} 0.11 \\ 0.11 \\ 0.11 \\ 0.11 \\ 0.11 \\ 0.11 \\ 0.11 \\ 0.11 \\ 0.11
  \end{array}$ & $\begin{array}{cc}-20.01 & 1.05\\ -14.97 & 1.05\\ -9.99 & 1.07\\ -5.02 & 1.05\\ 0.04 & 1.00\\ 4.97 & 1.88 \\ 10.00 & 0.53\\ 15.01 & 0.55\\ 20.00 & 0.57
  \end{array}$ & $\begin{array}
    {c} 0.11 \\ 0.11 \\ 0.11 \\ 0.11 \\ 0.11 \\ 0.11 \\ 0.12 \\ 0.11 \\ 0.11
  \end{array}$ & $\begin{array}
    {c} \wh\beta: 0.13\\ \scriptstyle{(-0.41,0.67)}\\
    \\
    \wh K: 11\\ \scriptstyle{(10,12)}\\
    \\
    \wh K_+: 9\\ \scriptstyle{(9,11)}
  \end{array}$\\
  \hline
\end{tabular}
\caption{{\footnotesize Static case. Results of a Monte Carlo exercise with $100$ iterations. Study of the impact when we increase the number of components. $\beta^{\star} = 0$, $K^{\star} = 9$, $T = 3$. The estimation are means across the $100$ Monte Carlo iterations of the posterior means. The credible interval (CI) for $\beta$ is the $95\%$ CI, and the $1^{st}$ and $3^{rd}$ quartiles for $\wh K$ and $\wh K_+$.}}\label{tab:3:static:MC}
\end{table}
}
\normalsize
Next, we have considered the dynamic case where $\gamma^{\st}$ is set equal to $0.1$. The results for this case are postponed to Appendix \ref{app:additional:simulations}.

\subsection{Impact of covariates on the group structure}\label{ss:sim:covariates}
Whether we include or not covariates in our model can affect the capability of our algorithm to detect the probabilistic model of the group structure depending on the strength of the omitted signal. We illustrate this fact with the following simulation. Suppose that data are generated according with
\begin{equation}\label{ss:sim:3}
  y_{it} = \beta^{\st} z_{it} + \alpha_i + u_{it},
\end{equation}
\noindent where $u_{it}\sim\mathcal{N}(0,\sigma_i^2)$, $(\alpha_i,\sigma_{i}^2)$ are drawn from a discrete distribution with atoms $\theta_{1,\alpha}^{\st} = -5$, $\theta_{2,\alpha}^{\st} = 0$, and $\theta_{3,\alpha}^{\st} = 5$ for $\alpha$, and $\theta_{1,\sigma^2}^{\st} = 0.1$, $\theta_{2,\sigma^2}^{\st} = 0.1$, and $\theta_{3,\sigma^2}^{\st} = 0.1$ for $\sigma^2$. The weights are $\bw= (0.45,0.5,0.05)$. The covariate $z_{it}$ is generated from a $\mathcal{N}(1,1)$.\\
\indent However, when we estimate the model we ignore $z_{it}$ and estimate the model without covariates, that is, we estimate the model $y_{it} = \alpha_i + \wtl{u}_{it}$, where $\wtl{u}_{it} = \beta^{\st} z_{it} + u_{it}$. We see that when the signal is very large, \textit{i.e.} $\beta^{\st} = 100$, it dominates the group structure so that for small $T$ we are not able to recovering the clustering structure and all the observations are boiled down in the same group. On the other hand, when the signal is smaller, \textit{i.e.} $\beta^{\st} = 10$, then we are able to recover the true number of groups for moderately large $T$ but the values of the atoms are inflated by the value of $\beta^{\st}$. The explanation for this is that the latent heterogeneity due to membership to different groups is blurred by a strong omitted signal. In fact, as this strong covariates are omitted they are in the error term $\wtl{u}_{it}$ making more difficult disentangle the clustering structure contained in $\alpha_i,\sigma_i^2$ is their variance is small compared to the omitted signal, that is, the signal-to-noise ratio for the omitted covariates is high.\\
\indent The results are reported in Table \ref{tab:1:omitt:MC} for $\beta^{\st} = 100$ (very strong signal) and in Table \ref{tab:2:omitt:MC} for $\beta^{\st} = 10$ (weaker signal). In the first case, the signal-to-noise ratio is $10,000/0.1 = 10^4$, while in the second case it is equal to $10^3$. Table $\ref{tab:2:omitt:MC}$ shows that for a signal-to-noise ratio equal to $10^3$ if we increase the time series from $T=3$ to $T=100$ we are able to recover the correct number of clusters. On the other hand, the atoms cannot be well estimated because they are not identified in this case. Table \ref{tab:3:omitt:MC} shows that as long as we include the previously omitted signal, estimation rapidly improves and we are able to recover the clustering structure (including the atoms).

{\footnotesize
\begin{table}[!h]
\centering
\begin{tabular}{|c|c|c|c||c|c|c||}
  \hline
  N & T & $\theta_{\alpha}^{\star},\theta_{\sigma^2}^{\star}$ & $w^{\star}$  & $\wh\theta$ & $\wh w_K$  & $\widehat{K} $, $\widehat{K}_+ $\\
  \hline
  \hline
  $50$ & $3$ & $\begin{array}{cc}-5 & 0.1\\ 0 & 0.1\\ 5 & 0.1\end{array}$ & $\begin{array}
    {c} 0.45 \\ 0.5 \\ 0.05 \end{array}$ & $\begin{array}{cc} 97.1 & 10283\end{array}$ & $\begin{array}
    {c} 1
  \end{array}$ & $\begin{array}
    {cl}
    \wh K: & \; 1\\ & \scriptstyle{(1,1)}\\
    \wh K_+: & \; 1\\ & \scriptstyle{(1,1)}
  \end{array}$\\
  \hline
  $50$ & $100$ & $\begin{array}{cc}-5 & 0.1\\ 0 & 0.1\\ 5 & 0.1\end{array}$ & $\begin{array}
    {c} 0.45 \\ 0.5 \\ 0.05 \end{array}$ & $\begin{array}{cc} 98.1 & 9990\end{array}$ & $\begin{array}
    {c} 1
  \end{array}$ & $\begin{array}
    {cl}
    \wh K: & \; 1\\ & \scriptstyle{(1,1)}\\
    \wh K_+: & \; 1\\ & \scriptstyle{(1,1)}
  \end{array}$\\
  \hline
  \hline
\end{tabular}
\caption{{\footnotesize Model \eqref{ss:sim:3} with $\beta^{\st} = 100$ and without $z_{it}$ in the estimtaion. Results of a Monte Carlo exercise with $100$ iterations. Study of the impact when we omit the explanatory variables. The estimation are means across the $100$ Monte Carlo iterations of the posterior means. }}\label{tab:1:omitt:MC}
\end{table}
}
\normalsize

{\footnotesize
\begin{table}[!h]
\centering
\begin{tabular}{|c|c|c|c||c|c|c||}
  \hline
  N & T & $\theta_{\alpha}^{\star},\theta_{\sigma^2}^{\star}$ & $w^{\star}$  & $\wh\theta$ & $\wh w_K$  & $\widehat{K} $, $\widehat{K}_+ $\\
  \hline
  \hline
  $50$ & $3$ & $\begin{array}{cc}-5 & 0.1\\ 0 & 0.1\\ 5 & 0.1\end{array}$ & $\begin{array}
    {c} 0.45 \\ 0.5 \\ 0.05 \end{array}$ & $\begin{array}{cc} 7.92 & 111\end{array}$ & $\begin{array}
    {c} 1
  \end{array}$ & $\begin{array}
    {cl}
    \wh K: & \; 1\\ & \scriptstyle{(1,1)}\\
    \wh K_+: & \; 1\\ & \scriptstyle{(1,1)}
  \end{array}$\\
  \hline
  $50$ & $100$ & $\begin{array}{cc}-5 & 0.1\\ 0 & 0.1\\ 5 & 0.1\end{array}$ & $\begin{array}
    {c} 0.45 \\ 0.5 \\ 0.05 \end{array}$ & $\begin{array}{cc} 5.04 & 99.9\\ 9.99 & 100.3 \\ 15.06 & 100 \end{array}$ & $\begin{array}
    {c} 0.45 \\ 0.47 \\ 0.07
  \end{array}$ & $\begin{array}
    {cl}
    \wh K: & \; 3\\ & \scriptstyle{(1,1)}\\
    \wh K_+: & \; 3\\ & \scriptstyle{(1,1)}
  \end{array}$\\
  \hline
  \hline
\end{tabular}
\caption{{\footnotesize Model \eqref{ss:sim:3} with $\beta^{\st} = 10$ and without $z_{it}$ in the estimation. Results of a Monte Carlo exercise with $100$ iterations. Study of the impact when we omit the explanatory variables. The estimation are means across the $100$ Monte Carlo iterations of the posterior means. }}\label{tab:2:omitt:MC}
\end{table}
}
\normalsize

{\footnotesize
\begin{table}[!h]
\centering
\begin{tabular}{|c|c|c|c||c|c|c||}
  \hline
  N,T & $\beta^*$ & $\theta_{\alpha}^{\star},\theta_{\sigma^2}^{\star}$ & $w^{\star}$  & $\wh\theta$ & $\wh w_K$  & $\widehat{K} $, $\widehat{K}_+ $\\
  \hline
  \hline
  $50$, $3$ & $100$ & $\begin{array}{cc}-5 & 0.1\\ 0 & 0.1\\ 5 & 0.1\end{array}$ & $\begin{array}
    {c} 0.45 \\ 0.5 \\ 0.05 \end{array}$ & $\begin{array}{cc} -2.93 & 2.84\\ 2.34 & 0.71 \end{array}$ & $\begin{array}
    {c} 0.67\\0.33
  \end{array}$ & $\begin{array}
    {cl}
    \wh K: & \; 2\\ & \scriptstyle{(1,1)}\\
    \wh K_+: & \; 2\\ & \scriptstyle{(1,1)}
  \end{array}$\\
  \hline
  $50$, $3$ & $10$ & $\begin{array}{cc}-5 & 0.1\\ 0 & 0.1\\ 5 & 0.1\end{array}$ & $\begin{array}
    {c} 0.45 \\ 0.5 \\ 0.05 \end{array}$ & $\begin{array}{cc} -5.00 & 0.11\\ 0.00 & 0.10 \\ 5.00 & 0.12 \end{array}$ & $\begin{array}
    {c} 0.44 \\ 0.49 \\ 0.07
  \end{array}$ & $\begin{array}
    {cl}
    \wh K: & \; 3\\ & \scriptstyle{(1,1)}\\
    \wh K_+: & \; 3\\ & \scriptstyle{(1,1)}
  \end{array}$\\
  \hline
  \hline
\end{tabular}
\caption{{\footnotesize Model \eqref{ss:sim:3} with $z_{it}$ in the estimation and different values of $\beta^{\st}$. Results of a Monte Carlo exercise with $100$ iterations. Study of the impact of the value of $\beta^{\st}$. The estimation are means across the $100$ Monte Carlo iterations of the posterior means. }}\label{tab:3:omitt:MC}
\end{table}
}
\normalsize

\section{Application: Income and Democracy }\label{s:Application}
We apply our procedure to analyse the statistical association between income and democracy across countries which is a cornerstone of modernization theory in political science and economics (\textit{e.g.} \cite{Lipset1959}, \cite{Rueschemeyer1992}, \cite{Barro1999}). This relationship is revisited in \cite{AcemougluEtAl2008} who show that once we control for factors that simultaneously affect both income and democracy, by including country fixed effects, this statistical association disappears. They use the Freedom House Political Rights Index as measure of democracy and the GDP per capita as a measure of income.\\
\indent More recently, \cite{BonhommeManresa2015} have analyzed this empirical question by arguing that countries can be grouped based on their level of democracy. They consider four groups: ``high-democracy'', ``low democracy'', ``early transition'' and ``low transition''. We refer to \cite[Section 4]{BonhommeManresa2015} for an explanation of these groups.\\
\indent In our study, instead of imposing a fixed number of groups, we treat this number as random and endow it with a prior according with our MFM modeling. The data that we use are taken from the replication files of \cite{BonhommeManresa2015} which in turn come from \cite{AcemougluEtAl2008} and we refer to these papers for a description of the dataset. The measure of income is the GDP per capita. The measure of democracy used is the Freedom House Political Rights Index constructed such that a country receives the highest score if political rights come closest to the ideals suggested by a checklist of questions. Using this index, \cite{AcemougluEtAl2008} have constructed five-year, ten-year, twenty-year, and annual panels. We try both five-years and annual panels and we use the sample period 1970-2000 for the five-year panel and the sample period 1975-2000 for the annual panel. We retain only the countries that have observations for all the years (or for all the 7 five-year periods) in this time span. For the five-year panel we have $N=92$ and $T=6$ while for the annual panel we have $N=97$ and $T=25$ (after loosing one period to account for the lagged variables).\\
\indent We start by estimating model \ref{eq:2} for $y_{i,t}$ given by the democracy measure, $h = 1$, and $z_{i,t-1}$ equal to the lagged log-GDP per capita. We are interested in understanding the effect of $\log(GDP)$ on democracy, which is given by the parameter $\beta$, and the implied cumulative income effect measured by $\beta/(1-\gamma)$. Table \ref{tab:1:application1:parameters} reports posterior mean estimates of $\gamma$, $\beta$ and $\beta/(1-\gamma)$ for different priors for $K$ and $v$. Table \ref{tab:1:application1:K} reports the posterior distribution of $K$ and $K_+$ for different priors on $K$ and on $v$. The following conclusions can be drawn. (1) As long as $K$ is random, and not fixed to a value, the estimate of the parameters and the posteriors of $K$ and $K_+$ are almost insensitive to variations in the prior of $K$ and of $v$ (in both the five-year and the annual panel). The distribution of $K$ and $K_+$ is highly concentrated on the value $1$. This important finding shows that there is no support in the data for more than one group. (2) A $10\%$ increase in income per capita is associated with a $10\%$ increase in the Freedom Hose index (for $5$-years panel) and a $2\%$ increase for annual panels. The implied cumulative income effect is about $0.2$ or $0.3$. The $95\%$-credible intervals for the corresponding parameters are tight and do not include the zero suggesting that there is an effect of income on democracy but that it is very small. The autoregressive parameter $\gamma$ is estimated at about $0.9$ in the annual panel and $0.6$ in the $5$-years panel indicating that there is a high degree of persistence in democracy. Our estimates for the $5$-years panel are similar to the ones obtained in \cite{BonhommeManresa2015} with one group. (3) When we use a degenerate prior for $K$ with a point mass on $K=10$, the estimates are higher: the posterior mean of $\beta$ is about $1$ and it is slightly smaller than $1$ for the parameter $\beta/(1-\gamma)$. The distribution of $K$ and $K_+$ is still concentrated on the value $1$ but with a smaller mass than in the random-$K$ case.
{\footnotesize
\begin{table}[!h]
\centering
\begin{tabular}{|c|c||c|c|c||c|c|c||}
  \hline
  \multirow{2}{*}{$\Pi(K)$} & \multirow{2}{*}{$\Pi(v)$} & \multicolumn{3}{c||}{$5$-year panel} & \multicolumn{3}{c||}{Annual panel}  \\
  \cline{3-8}
  & & $\gamma$ & $\beta$ & $\beta/(1-\gamma)$ & $\gamma$ & $\beta$ & $\beta/(1-\gamma)$\\
  \hline
  \hline
  {\tiny $BNB(1,4,3)$} & $\delta_1$ & $\begin{array}
    {c} 0.60\\ \scriptstyle{(0.20,0.72)}
  \end{array}$ & $\begin{array}
    {c} 0.11\\ \scriptstyle{(0.05,0.28)}
  \end{array}$ & $\begin{array}
    {c} 0.25\\ \scriptstyle{(0.18,0.41)}
  \end{array}$ & $\begin{array}
    {c} 0.92\\ \scriptstyle{(0.83,0.94)}
  \end{array}$ & $\begin{array}
    {c} 0.02\\ \scriptstyle{(0.01,0.06)}
  \end{array}$ & $\begin{array}
    {c} 0.21\\ \scriptstyle{(0.15,0.39)}
  \end{array}$ \\
  \hline
  {\tiny $\mathcal{U}\{0,4\}$} & $\delta_1$ & $\begin{array}
    {c} 0.59\\ \scriptstyle{(0.26,0.73)}
  \end{array}$ & $\begin{array}
    {c} 0.11\\ \scriptstyle{(0.05,0.26)}
  \end{array}$ & $\begin{array}
    {c} 0.26\\ \scriptstyle{(0.16,0.39)}
  \end{array}$ & $\begin{array}
    {c} 0.91\\ \scriptstyle{(0.77,0.94)}
  \end{array}$ & $\begin{array}
    {c} 0.02\\ \scriptstyle{(0.01,0.10)}
  \end{array}$ & $\begin{array}
    {c} 0.21\\ \scriptstyle{(0.15,0.44)}
  \end{array}$ \\
  \hline
  $\delta_{10}$ & $\delta_1$ & $\begin{array}
    {c} -0.60\\ \scriptstyle{(-0.98,0.02)}
  \end{array}$ & $\begin{array}
    {c} 1.04\\ \scriptstyle{(0.44,3.31)}
  \end{array}$ & $\begin{array}
    {c} 0.62\\ \scriptstyle{(0.35,1.71)}
  \end{array}$ & $\begin{array}
    {c} -0.44\\ \scriptstyle{(-0.98,0.25)}
  \end{array}$ & $\begin{array}
    {c} 1.33\\ \scriptstyle{(0.39,8.51)}
  \end{array}$ & $\begin{array}
    {c} 0.81\\ \scriptstyle{(0.36,4.33)}
  \end{array}$ \\
  \hline
  $\delta_{10}$ & {\tiny $\mathcal{G}a(1,20)$} & $\begin{array}
    {c} -0.28\\ \scriptstyle{(-0.98,0.61)}
  \end{array}$ & $\begin{array}
    {c} 0.73\\ \scriptstyle{(0.11,2.80)}
  \end{array}$ & $\begin{array}
    {c} 0.52\\ \scriptstyle{(0.28,1.45)}
  \end{array}$ & $\begin{array}
    {c} 0.39\\ \scriptstyle{(-0.98,0.34)}
  \end{array}$ & $\begin{array}
    {c} 1.25\\ \scriptstyle{(0.34,8.32)}
  \end{array}$ & $\begin{array}
    {c} 0.79\\ \scriptstyle{(0.36,4.25)}
  \end{array}$ \\
  \hline
  {\tiny $\mathcal{G}eom(0.2)$} & $\delta_6$ & $\begin{array}
    {c} 0.66\\ \scriptstyle{(0.59,0.72)}
  \end{array}$ & $\begin{array}
    {c} 0.08\\ \scriptstyle{(0.06,0.10)}
  \end{array}$ & $\begin{array}
    {c} 0.23\\ \scriptstyle{(0.18,0.28)}
  \end{array}$ & $\begin{array}
    {c} 0.92\\ \scriptstyle{(0.90,0.94)}
  \end{array}$ & $\begin{array}
    {c} 0.02\\ \scriptstyle{(0.01,0.02)}
  \end{array}$ & $\begin{array}
    {c} 0.20\\ \scriptstyle{(0.15,0.27)}
  \end{array}$ \\
  \hline
\end{tabular}
\caption{{\footnotesize Income and Democracy. Static MFM with atoms independent of $\bZ$. Mean estimation of the parameters. Results for different priors on $K$ and $v$.}}\label{tab:1:application1:parameters}
\end{table}
}
\normalsize

{\footnotesize
\begin{table}[!h]
\centering
\begin{tabular}{|c|c||c|c|c|c||c|c|c|c||}
  \hline
  \multirow{2}{*}{$\Pi(K)$} & \multirow{2}{*}{$\Pi(v)$} & \multicolumn{4}{c||}{$5$-year panel} & \multicolumn{4}{c||}{Annual panel}  \\
  \cline{3-10}
  & & \multicolumn{4}{c||}{$\Pi(K = k|\by,\bZ,\by_0)$} & \multicolumn{4}{c||}{$\Pi(K = k|\by,\bZ,\by_0)$}\\
  \cline{3-10}
  & & $k=1$ & $k=2$ & $k=3$ & $k=4$ & $k=1$ & $k=2$ & $k=3$ & $k=4$ \\
  \hline
  \hline
  {\tiny $BNB(1,4,3)$} & $\delta_1$ & $0.96$ & $0.04$ & $0.00$ & $0.00$ & $0.99$ & $0.01$ & $0.0001$ & $0.0001$\\
  \hline
  {\tiny $\mathcal{U}\{0,4\}$} & $\delta_1$ & $0.92$ & $0.07$ & $0.01$ & $0.00$ & $0.97$ & $0.03$ & $0.0012$ & $0$\\
  \hline
  $\delta_{10}$ & $\delta_1$ & $0$ & $0$ & $0$ & $0$ & $0$ & $0$ & $0$ & $0$\\
  \hline
  $\delta_{10}$ & $\mathcal{G}a(1,20)$ & $0$ & $0$ & $0$ & $0$ & $0$ & $0$ & $0$ & $0$\\
  \hline
  {\tiny $\mathcal{G}eom(0.2)$} & $\delta_6$ & $1$ & $0$ & $0$ & $0$ & $1$ & $0$ & $0$ & $0$\\
  \hline
  \hline
  & & \multicolumn{4}{c||}{$\Pi(K_+ = k|\by,\bZ,\by_0)$} & \multicolumn{4}{c}{$\Pi(K_+ = k|\by,\bZ,\by_0)$}\\
  \cline{3-10}
  & & $k=1$ & $k=2$ & $k=3$ & $k=4$ & $k=1$ & $k=2$ & $k=3$ & $k=4$ \\
  \hline
  \hline
  {\tiny $BNB(1,4,3)$} & v = 1 & $0.97$ & $0.03$ & $0.00$ & $0$ & $0.99$ & $0.01$ & $0.00$ & $0.00$\\
  \hline
  {\tiny $\mathcal{U}\{0,4\}$} & v = 1 & $0.94$ & $0.05$ & $0.00$ & $0$ & $0.99$ & $0.0114$ & $0$ & $0$\\
  \hline
  $\delta_{10}$ & $v=1$ & $0.68$ & $0.32$ & $0.00$ & $0$ & $0.66$ & $0.32$ & $0.02$ & $0.0004$\\
  \hline
  $\delta_{10}$ & $\mathcal{G}a(1,20)$ & $0.71$ & $0.29$ & $0.01$ & $.00$ & $0.66$ & $0.34$ & $0.0029$ & $0$\\
  \hline
  {\tiny $\mathcal{G}eom(0.2)$} & $\delta_6$ & $0.9999$ & $0$ & $0$ & $0$ & $1$ & $0$ & $0$ & $0$\\
  \hline
\end{tabular}
\caption{{\footnotesize Income and Democracy. Static MFM. Posterior distribution of $K$ and $K_+$. Results for different priors on $K$ and $v$.}}\label{tab:1:application1:K}
\end{table}
}
\normalsize


\subsection{Additional controls}
We have extended our empirical analysis to control for the following additional covariates: education, log-population size, percent population age for the following age groups: $0-15$, $15-30$, $30-45$, $45-60$, $60-$, and median age in the population. We use either a Poisson prior or a Beta-Negative-Binomial for the unknown number of groups: $K-1\sim \mathcal{P}oi(9)$ or $K-1 \sim BNB(1,4,3)$. \\
\indent The first striking result is that now we do not detect any causal effect of income on democracy. The second striking consequence of adding controls is that now we detect four clusters in the latent variables, which indicates that the fact that only one group was detected when controls were omitted was due to the omission of a strong signal that was blurring the clustering structure. The added controls explain a large part of the heterogeneity. The heterogeneity in the residuals when we add controls is therefore smaller in absolute value than the heterogeneity in the residuals obtained by accounting only for lagged democracy and GDP-per capita. The probabilistic structure of the residual heterogeneity when we add controls is well fitted by a mixing distribution with more than one components. On the other hand, without controls the probabilistic structure of the residual heterogeneity is well fitted by a mixing distribution with only one component. The fact that the detected number of components of the mixing distribution changes depending on the explanatory variables can be understood as follows: when part of the observed heterogeneity is omitted -- instead of accounted for explicitly as we do when we add controls -- it is more difficult to recover the mixing distribution of the unobserved heterogeneity because the signal-to-noise ratio is very high. This is in line with what we have illustrated in our numerical exercise in Section \ref{ss:sim:covariates}. In this case, we can see from the third column of Tables \ref{tab:application1:atoms}-\ref{tab:application2:atoms} that the noise (as captured by $\theta_{k,\sigma^2}$) is very small.\\
\indent To get a better insight we report in Figure \ref{fig:application:histograms} in the Appendix the histograms of: the data $y_{i,t}$ (Panel (a)), the residuals from model \ref{eq:2} with $z_{i,t-1}$ equal to the lagged log-GDP per capita (Panel (b)), the residuals from model \ref{eq:2} with $z_{i,t-1}$ containing the lagged log-GDP per capita and $\log(population\; size)$ (Panel (c)).\\

%
\indent The results of our estimation procedure are reported in Tables \ref{tab:2:application0:parameters} - \ref{tab:application2:atoms}. Each pair of tables refers to the two priors considered. Each row of the four tables refers to a different set of controls included in the regression model. Table \ref{tab:2:application0:parameters}- \ref{tab:2:application1:parameters} show that the effect of income on democracy is estimated to be almost zero in all the configurations considered. This result is in line with \cite{AcemougluEtAl2008} and indicates that there is no evidence for a strong causal effect of income on democracy after controlling for additional covariates and for unobserved heterogeneity. The fact that in the analysis without additional covariates we were founding a slightly positive $\beta$ was due to the omitted controls.\\
\indent In terms of probabilistic structure of the unobserved heterogeneity, we find four non-empty components but one of these components is characterized by a variance parameter $\theta_{k,\sigma^2}$ almost equal to zero meaning that this component is characterized by a Dirac mass at $\theta_{k,\alpha}$. We report these results as well as the value of the atoms in Tables \ref{tab:application1:atoms} and \ref{tab:application2:atoms}. The atoms are very similar for all the configurations considered.

{\footnotesize
\begin{table}[!h]
\centering
\begin{tabular}{|l|c|c||c|c|c||c|c|c||}
  \hline
  covariates & $\gamma$ & $\beta$ & $\beta/(1-\gamma)$ & $\wh{K}$ & $\wh{K}_+$ \\
  \hline
  \hline
  all & $\begin{array}
    {c} 0.13\\ \scriptstyle{(0.13,0.13)}
  \end{array}$ & $\begin{array}
    {c} -2.94e-07\\ \scriptstyle{(-1.93e-07,-1.11e-07)}
  \end{array}$ & $\begin{array}
    {c} -2.21e-07\\ \scriptstyle{(-3.37e-07,-1.27e-07)}
  \end{array}$ & $4$ & $4$\\
  \hline
  age-4 & $\begin{array}
    {c} 0.129\\ \scriptstyle{(0.128,0.130)}
  \end{array}$ & $\begin{array}
    {c} 9.88e-05\\ \scriptstyle{(6.10e-05,1.84e-04)}
  \end{array}$ & $\begin{array}
    {c} 1.13e-04\\ \scriptstyle{(7.01e-05,2.12e-04)}
  \end{array}$ & $4$ & $4$\\
  \hline
  ed-lpop & $\begin{array}
    {c} 0.13\\ \scriptstyle{(0.12,0.13)}
  \end{array}$ & $\begin{array}
    {c} 3.08e-06\\ \scriptstyle{(2.34e-08,1.38e-03)}
  \end{array}$ & $\begin{array}
    {c} 3.54e-06\\ \scriptstyle{(2.69e-08,1.57e-03)}
  \end{array}$ & $4$ & $4$\\
  \hline
  ed & $\begin{array}
    {c} 0.1496\\ \scriptstyle{(0.1209,0.1497)}
  \end{array}$ & $\begin{array}
    {c} 7.04e-06\\ \scriptstyle{(2.04e-08,3.61e-03)}
  \end{array}$ & $\begin{array}
    {c} 8.28e-06\\ \scriptstyle{(2.40e-08,4.11e-03)}
  \end{array}$ & $4$ & $4$\\
  \hline
  pop & $\begin{array}
    {c} 0.1252\\ \scriptstyle{(0.1036,0.1274)}
  \end{array}$ & $\begin{array}
    {c} 2.42e-04\\ \scriptstyle{(2.89e-05,2.92e-03)}
  \end{array}$ & $\begin{array}
    {c} 2.77e-04\\ \scriptstyle{(3.32e-05,3.26e-03)}
  \end{array}$ & $5$ & $5$\\
  \hline
\end{tabular}
\caption{{\footnotesize Income and Democracy. Static MFM and $K-1\sim\mathcal{P}oi(9)$ and $\Pi(v)= \delta_1(v)$. Estimation for different controls. ``age-4'' means age group percentages (four categories) in the population plus the median age in the population; ``ed-lpop'' means education and $\log(population\; size)$; ``ed'' means education; ``pop'' means $\log(population\;size)$.}}\label{tab:2:application0:parameters}
\end{table}

{\footnotesize
\begin{table}[!h]
\centering
\begin{tabular}{|l|c|c||c|c|c||c|c|c||}
  \hline
  covariates & $\gamma$ & $\beta$ & $\beta/(1-\gamma)$ & $\wh{K}$ & $\wh{K}_+$ \\
  \hline
  \hline
  all & $\begin{array}
    {c} 0.1275\\ \scriptstyle{(0.1274,0.1276)}
  \end{array}$ & $\begin{array}
    {c} -1.24e-05\\ \scriptstyle{(-1.96e-05,-8.01e-06)}
  \end{array}$ & $\begin{array}
    {c} -1.42e-05\\ \scriptstyle{(-2.25e-05,-9.18e-06)}
  \end{array}$ & $4$ & $4$\\
  \hline
  age-4 & $\begin{array}
    {c} 0.12\\ \scriptstyle{(0.12,0.13)}
  \end{array}$ & $\begin{array}
    {c} -5.96e-05\\ \scriptstyle{(-1.08e-04,-3.36e-05)}
  \end{array}$ & $\begin{array}
    {c} -6.76e-05\\ \scriptstyle{(-1.23e-04,-3.81e-05)}
  \end{array}$ & $4$ & $4$\\
  \hline
  ed-lpop & $\begin{array}
    {c} 0.15\\ \scriptstyle{(0.146,0.154)}
  \end{array}$ & $\begin{array}
    {c} 3.20e-06\\ \scriptstyle{(1.61e-08,1.01e-03)}
  \end{array}$ & $\begin{array}
    {c} 3.78e-06\\ \scriptstyle{(1.90e-08,1.19e-03)}
  \end{array}$ & $4$ & $4$\\
  \hline
  ed & $\begin{array}
    {c} 0.1593\\ \scriptstyle{(0.136,0.1594)}
  \end{array}$ & $\begin{array}
    {c} 1.00e-06\\ \scriptstyle{(7.45e-08,2.47e-03)}
  \end{array}$ & $\begin{array}
    {c} 1.19e-05\\ \scriptstyle{(8.87e-08,2.85e-03)}
  \end{array}$ & $4$ & $4$\\
  \hline
  pop & $\begin{array}
    {c} 0.1332\\ \scriptstyle{(0.1066,0.1332)}
  \end{array}$ & $\begin{array}
    {c} 2.01e-06\\ \scriptstyle{(2.52e-09,2.54e-03)}
  \end{array}$ & $\begin{array}
    {c} 2.32e-06\\ \scriptstyle{(2.90e-09,2.84e-03)}
  \end{array}$ & $4$ & $4$\\
  \hline
\end{tabular}
\caption{{\footnotesize Income and Democracy.  Static MFM and $K-1\sim BNB(1,4,3)$ and $\Pi(v)= \delta_1(v)$. Estimation for different controls. ``age-4'' means age group percentages (four categories) in the population plus the median age in the population; ``ed-lpop'' means education and $\log(population\; size)$; ``ed'' means education; ``pop'' means $\log(population\;size)$.}}\label{tab:2:application1:parameters}
\end{table}

{\footnotesize
\begin{table}[!h]
\centering
\begin{tabular}{|l||c|c|c||c||}
  \hline
  \scriptsize{covariates} & $\wh\theta_{\alpha}$ & $\wh\theta_{\sigma^2}$ & $\wh w_{K_+}$ & $\wh{K}_+$ \\
  \hline
  \hline
  all & $\begin{array}
    {cc} 0.17, & 0.79,\\ 0.87, & 0.47
  \end{array}$
  & $\begin{array}
    {cc} 0.02, & 0.01,\\ 1.39e-15, & 0.07
  \end{array}$ & $\begin{array}
    {cc} 0.19, & 0.12, \\ 0.20, & 0.48
  \end{array}$ & \scriptsize{$\Pi(K_+ = 4|\by,\bZ,\by_0) = 1$}\\
  \hline
  age-4 & $\begin{array}
    {cc} 0.79, & 0.87,\\ 0.17, & 0.47
  \end{array}$
  & $\begin{array}
    {cc} 0.01, & 4.90e-10,\\ 0.02, & 0.07
  \end{array}$ & $\begin{array}
    {cc} 0.12, & 0.20, \\ 0.19, & 0.48
  \end{array}$ & \scriptsize{$\Pi(K_+ = 4|\by,\bZ,\by_0) = 0.97$}\\
  \hline
  ed-lpop & $\begin{array}
    {cc} 0.17, & 0.79,\\ 0.87, & 0.47
  \end{array}$ & $\begin{array}
    {cc} 0.02, & 0.01,\\ 1.45e-08, & 0.07
  \end{array}$ & $\begin{array}
    {cc} 0.19, & 0.12,\\ 0.20, & 0.48
  \end{array}$ & \scriptsize{$\Pi(K_+ = 4|\by,\bZ,\by_0) = 1$}\\
  \hline
  ed & $\begin{array}
    {cc} 0.46, & 0.85,\\ 0.16, & 0.77
  \end{array}$ & $\begin{array}
    {cc} 0.07, & 3.83e-08,\\ 0.02, & 0.01
  \end{array}$ & $\begin{array}
    {cc} 0.49, & 0.20,\\ 0.18, & 0.12
  \end{array}$ & \scriptsize{$\Pi(K_+ = 4|\by,\bZ,\by_0) = 0.96$}\\
  \hline
  pop & $\begin{array}
    {cc} 0.16, & 0.47,\\ 0.87, & 0.79
  \end{array}$ & $\begin{array}
    {cc} 0.02, & 0.08,\\ 7.16e-08, & 0.12
  \end{array}$ & $\begin{array}
    {cc} 0.19, & 0.48,\\ 0.21, & 0.12
  \end{array}$ & \scriptsize{$\Pi(K_+ = 5|\by,\bZ,\by_0) = 0.93$}\\
  \hline
  \scriptsize{no controls} & $-4.43$ & $0.59$ & $1$ & \scriptsize{$\Pi(K_+ = 1|\by,\bZ,\by_0) = 0.71$}\\
  \hline
\end{tabular}
\caption{{\footnotesize Income and Democracy. Static MFM and $K-1\sim\mathcal{P}oi(9)$. Estimation for different controls. ``age-4'' means age group percentages (four categories) in the population plus the median age in the population; ``ed-lpop'' means education and $\log(population\; size)$; ``ed'' means education; ``pop'' means $\log(population\;size)$. Notice that in ``pop'' there are $5$ clusters but two have degenerate distributions at $0.87$.}}\label{tab:application1:atoms}
\end{table}

{\footnotesize
\begin{table}[!h]
\centering
\begin{tabular}{|l||c|c|c||c||}
  \hline
  \scriptsize{covariates} & $\wh\theta_{\alpha}$ & $\wh\theta_{\sigma^2}$ & $\wh w_{K_+}$ & $\wh{K}_+$ \\
  \hline
  \hline
  all & $\begin{array}
    {cc} 0.17, & 0.79,\\ 0.87, & 0.47
  \end{array}$
  & $\begin{array}
    {cc} 0.02, & 0.01,\\ 6.07e-12, & 0.07
  \end{array}$ & $\begin{array}
    {cc} 0.19, & 0.12, \\ 0.21, & 0.48
  \end{array}$ & \scriptsize{$\Pi(K_+ = 4|\by,\bZ,\by_0) = 1$}\\
  \hline
  age-4 & $\begin{array}
    {cc} 0.17, & 0.80,\\ 0.88, & 0.48
  \end{array}$
  & $\begin{array}
    {cc} 0.02, & 0.01,\\ 1.68e-10, & 0.08
  \end{array}$ & $\begin{array}
    {cc} 0.20, & 0.12, \\ 0.20, & 0.48\\
    0.03 &
  \end{array}$ & \scriptsize{$\Pi(K_+ = 4|\by,\bZ,\by_0) = 1$}\\
  \hline
  ed-lpop & $\begin{array}
    {cc} 0.16, & 0.77,\\ 0.85, & 0.45
  \end{array}$ & $\begin{array}
    {cc} 0.02, & 0.01,\\ 7.60e-09, & 0.07
  \end{array}$ & $\begin{array}
    {cc} 0.17, & 0.12,\\ 0.20, & 0.50
  \end{array}$ & \scriptsize{$\Pi(K_+ = 4|\by,\bZ,\by_0) = 1$}\\
  \hline
  ed & $\begin{array}
    {cc} 0.16, & 0.76,\\ 0.84, & 0.45
  \end{array}$ & $\begin{array}
    {cc} 0.02, & 0.01,\\ 4.66e-08, & 0.07
  \end{array}$ & $\begin{array}
    {cc} 0.17, & 0.12,\\ 0.21, & 0.50
  \end{array}$ & \scriptsize{$\Pi(K_+ = 4|\by,\bZ,\by_0) = 1$}\\
  \hline
  pop & $\begin{array}
    {cc} 0.17, & 0.79,\\ 0.87, & 0.47
  \end{array}$ & $\begin{array}
    {cc} 0.02, & 0.01,\\ 5.32e-08, & 0.07
  \end{array}$ & $\begin{array}
    {cc} 0.18, & 0.12,\\ 0.20, & 0.49
  \end{array}$ & \scriptsize{$\Pi(K_+ = 4|\by,\bZ,\by_0) = 1$}\\
  \hline
  \scriptsize{no controls} & $-0.55$ & $0.04$ & $1$ & \scriptsize{$\Pi(K_+ = 1|\by,\bZ,\by_0) = 0.96$}\\
  \hline
\end{tabular}
\caption{{\footnotesize Income and Democracy. Static MFM and $K-1\sim BNB(1,4,3)$. Estimation for different controls. ``age-4'' means age group percentages (four categories) in the population plus the median age in the population; ``ed-lpop'' means education and $\log(population\; size)$; ``ed'' means education; ``pop'' means $\log(population\;size)$.}}\label{tab:application2:atoms}
\end{table}

\normalsize

\section{Conclusions}\label{s:conclusions}
This paper proposes a structural framework for modeling unobserved heterogeneity in dynamic panel data through a mixture of finite mixtures (MFMs) specification. Our approach jointly estimates the regression parameters and the clustering structure, without fixing the number of groups in advance.\\
\indent There are five main contributions. First, we provide a probabilistic model of clustering in panel data models, moving beyond approaches that use groups as a tool to approximate unobserved heterogeneity. Second, we study the prior on the number of clusters and the sensitivity of the results to it, clarifying the distinction between the true number of groups and those effectively represented in finite samples. Third, we establish asymptotic guarantees, showing that the posterior distribution of the mixing measure contracts around the truth at near-parametric rates. Fourth, we extend the Telescoping Sampler of \cite{FruwirthSchnatter2021} to panel settings, yielding an efficient algorithm for posterior inference. Fifth, we show that the ability of recovering the clustering structure depends on the signal-to-noise ratio and that if a strong signal is omitted, then this can heavily impact the ability to detect a group structure in finite samples. In the latter case, all the individuals are put in the same group simply because we have omitted important variables from the model.

Monte Carlo simulations confirm that the method recovers the clustering structure well when groups are separated, and remains reliable for the regression parameters even in more difficult cases. Importantly, inference for the common regression parameters remains accurate in all cases. In the application to the income-democracy relationship, we find no evidence of multiple clusters when controls are omitted. When we account for important controls then, we indeed find that the data support four latent groups as suggested by previous literature.

Overall, our results show that structural modeling of latent clustering in panels is both feasible and informative, offering a new perspective on the analysis of heterogeneous economic agents.\\
\paragraph{Acknowledgments.} The authors thank Guillaume Kon Kam King and Daria Bystrova for pointing out the Telescoping sampler of \cite{FruwirthSchnatter2021}, Stephane Bonhomme, and the participants to the following conferences and seminars: 2024 Women in Econometrics Conference - University of Bologna, LPSM Sorbonne Universit\'{e}, Nuffield College, CORE Louvain-La-Neuve, Journ\'{e}es de la statistique ENSAE-ENSAI-INSEE, BNP14, German Statistical Week 2025. Jean-Pierre Florens acknowledges funding from the French National Research Agency (ANR) under Investment for the future (Investissement d'Avenir) Grant ANR-17-EURE-0010. Anna Simoni gratefully acknowledges financial support from \href{https://www.hi-paris.fr/}{Hi!Paris}, and ANR-21-CE26-0003.
\paragraph{Online Appendix.} It contains: the proofs, and details for the implementation of the telescoping sampler for dynamic panel data. 
\appendix
%
%
%
%
\section{Additional simulation results}\label{app:additional:simulations}
Here, we show the results of our numerical experiments for the dynamic case where $\gamma^{\st}$ is set equal to $0.1$. We have tried different values for $K^{\st}$, $\bw^{\st}$, $\btheta^{\st}$, $N$ and $T$. The results are reported in Tables \ref{tab:1:dynamic:MC}-\ref{tab:3:dynamic:MC}. Table \ref{tab:1:dynamic:MC} presents the result of our procedure when we vary $N$ and $T$ in a situation where $K^{\st} = 3$ and there is enough variation in the components of $\theta_{\alpha}$ (while all the components of $\theta_{\sigma^2}$ are set equal to $1$). The results are very good even for small value of $N$ and $T$ (that is, $N=50$ and $T = 3$) and so there is no gain in increasing $N,T$.\\
\indent In Table \ref{tab:2:dynamic:MC} we analyse the impact of having some atoms very similar across components. For $K^{\st} = 3$, we set $\theta_{2,\alpha} = 0$ and $\theta_{3,\alpha} = 0.5$. As in the static case, we see that estimation of the common parameters $\beta^{\st}$ and $\gamma^{\st}$ is very good even for small values of $N$ and $T$. Instead, in order to recover the clustering structure we need a larger than $3$ time dimension if the cross-section dimension $N$ is small. For instance, with $N=50$ and $T=100$ we estimate the clustering structure very precisely.\\
\indent Table \ref{tab:3:dynamic:MC} considers the effect of increasing the number of clusters on the estimation performance of our method. We consider $K^{\st}=9$ components with the first components of the atoms well separated and the second component being the same for all the components. When $N=50$, the mean across the $100$ MC iterations is found to be $\wh{K}_+ = 4$. In this case with $K^{\st}=9$ there is no MC iteration with a number of clusters equal to $4$. Therefore, we have estimated the atoms and the corresponding weights by averaging over the MC iterations with exactly $K^*=9$ clusters. These ones represents only $16$ MC iterations over $100$. Instead, by averaging over the MC iterations with a number of clusters equal to the most frequent number of estimated clusters across the $100$ MC iterations (which is $1$ in our exercise) we get an average estimator for $\btheta_1$ equal to $(-0.00, 1.77)$ and a corresponding weight equal to $1$. This anomaly disappears when $N$ increases, for instance $N=500$.

{\footnotesize
\begin{table}[!h]
\centering
\begin{tabular}{|c|c|c|c||c|c|c|c||}
  \hline
  N & T & $\theta_{\alpha}^{\star},\theta_{\sigma^2}^{\star}$ & $w^{\star}$  & $\wh\theta$ & $\wh w_K$  & $\wh\beta$, $\wh\gamma$ & $\widehat{K} $, $\widehat{K}_+ $\\
  \hline
  \hline
  $50$ & $3$ & $\begin{array}{cc}-5 & 1\\ 0 & 1\\ 5 & 1\end{array}$ & $\begin{array}
    {c} 1/3 \\ 1/3 \\ 1/3 \end{array}$ & $\begin{array}{cc}-4.50 & 1.24\\ -0.02 & 1.21\\ 5.13 & 1.19\end{array}$ & $\begin{array}
    {c} 0.34 \\ 0.35 \\ 0.32
  \end{array}$ & $\begin{array}
    {c} \wh\beta: -0.03\\ \scriptstyle{(-0.29,0.26)}\\
    \wh\gamma: 0.17\\ \scriptstyle{(-0.06,0.34)}
  \end{array}$ & $\begin{array}
    {cl}
    \wh K: & \; 3\\ & \scriptstyle{(3,3)}\\
    \wh K_+: & \; 3\\ & \scriptstyle{(3,3)}
  \end{array}$\\
  \hline
  $50$ & $30$ & $\begin{array}{cc}-5 & 1\\ 0 & 1\\ 5 & 1\end{array}$ & $\begin{array}
    {c} 1/3 \\ 1/3 \\ 1/3 \end{array}$ & $\begin{array}{cc}-5.04 & 1.02\\ -0.01 & 1.02\\ 5.03 & 1.01\end{array}$ & $\begin{array}
    {c} 0.33 \\ 0.34 \\ 0.33
  \end{array}$ & $\begin{array}
    {c} \wh\beta: -0.003\\ \scriptstyle{(-0.06,0.05)}\\
    \wh\gamma: 0.09\\ \scriptstyle{(0.04,0.14)}
  \end{array}$ & $\begin{array}
    {cl}
    \wh K: & \; 3\\ & \scriptstyle{(3,3)}\\
    \wh K_+: & \; 3\\ & \scriptstyle{(3,3)}
  \end{array}$\\
  \hline
  $100$ & $3$ & $\begin{array}{cc}-5 & 1\\ 0 & 1\\ 5 & 1\end{array}$ & $\begin{array}
    {c} 1/3 \\ 1/3 \\ 1/3
  \end{array}$ & $\begin{array}{cc}-4.97 & 1.11\\ 0.00 & 1.12 \\ 4.98 & 1.11\end{array}$ & $\begin{array}
    {c} 0.33\\ 0.33\\ 0.34
  \end{array}$ & $\begin{array}
    {c} \wh{\beta}: 0.04\\ \scriptstyle{(-0.11,0.20)}\\
    \wh{\gamma}: 0.10\\
    \scriptstyle{(-0.02,0.22)}
  \end{array}$ & $\begin{array}{cl} \wh{K}: & \; 3\\ & \scriptstyle{(3,3)}\\
  \wh{K}_+: & \; 3 \\ & \scriptstyle{(3,3)}\end{array}$ \\
  \hline
  $500$ & $3$ & $\begin{array}{cc}-5 & 1\\ 0 & 1\\ 5 & 1\end{array}$ & $\begin{array}
    {c} 1/3 \\ 1/3 \\ 1/3
  \end{array}$ & $\begin{array}{cc}-5.03 & 1.03\\ -0.00 & 1.03 \\ 5.02 & 1.04\end{array}$ & $\begin{array}
    {c} 0.33\\ 0.34\\ 0.33
  \end{array}$ & $\begin{array}
    {c} \wh{\beta}: -0.01\\ \scriptstyle{(-0.07,0.05)}\\
    \wh{\gamma}: 0.10\\
    \scriptstyle{(0.04,0.15)}
  \end{array}$ & $\begin{array}{cl} \wh{K}: & \; 3\\ & \scriptstyle{(3,3)}\\
  \wh{K}_+: & \; 3 \\ & \scriptstyle{(3,3)}\end{array}$ \\
  \hline
\end{tabular}
\caption{{\footnotesize Dynamic case. Results of a Monte Carlo exercise with $100$ iterations. Study of the impact when we increase $T$ and $N$. $\beta^{\star} = 0$, $\gamma^{\st} = 0.1$, $K^{\star} = 3$. The estimation are means across the $100$ Monte Carlo iterations of the posterior means. The credible intervals (CI) for $\beta$ and $\gamma$ are the $95\%$ CI, and the $1^{st}$ and $3^{rd}$ quartiles for $\wh K$ and $\wh K_+$.}}\label{tab:1:dynamic:MC}
\end{table}
}
\normalsize

{\footnotesize
\begin{table}[!h]
\centering
\begin{tabular}{|c|c|c|c||c|c|c|c||}
  \hline
  N & T & $\theta_{\alpha}^{\star},\theta_{\sigma^2}^{\star}$ & $w^{\star}$  & $\wh\theta$ & $\wh w_K$  & $\wh\beta$, $\wh\gamma$ & $\widehat{K} $, $\widehat{K}_+ $\\
  \hline
  \hline
  $50$ & $3$ & $\begin{array}{cc}-5 & 1\\ 0 & 1\\ 0.5 & 1\end{array}$ & $\begin{array}
    {c} 0.45 \\ 0.5 \\ 0.05 \end{array}$ & $\begin{array}{cc}-4.76 & 1.11\\ 0.05 & 1.13\end{array}$ & $\begin{array}
    {c} 0.49 \\ 0.53
  \end{array}$ & $\begin{array}
    {c} \wh\beta: 0.01\\ \scriptstyle{(-0.20,0.22)}\\
    \wh\gamma: 0.12\\ \scriptstyle{(-0.08,0.30)}
  \end{array}$ & $\begin{array}
    {cl}
    \wh K: & \; 2\\ & \scriptstyle{(2,2)}\\
    \wh K_+: & \; 2\\ & \scriptstyle{(2,2)}
  \end{array}$\\
  \hline
  $500$ & $3$ & $\begin{array}{cc}-5 & 1\\ 0 & 1\\ 0.5 & 1\end{array}$ & $\begin{array}
    {c} 0.45 \\ 0.5 \\ 0.05 \end{array}$ & $\begin{array}{cc} -4.95 & 1.01\\ 0.05 & 1.03 \end{array}$ & $\begin{array}
    {c} 0.45 \\0.55
  \end{array}$ & $\begin{array}
    {c} \wh\beta: 0.00\\ \scriptstyle{(-0.05,0.05)}\\
    \wh\gamma: 0.11\\ \scriptstyle{(0.05,0.16)}
  \end{array}$ & $\begin{array}
    {cl}
    \wh K: & \; 2\\ & \scriptstyle{(2,2)}\\
    \wh K_+: & \; 2\\ & \scriptstyle{(2,2)}
  \end{array}$\\
  \hline
  $50$ & $100$ & $\begin{array}{cc}-5 & 1\\ 0 & 1\\ 0.5 & 1\end{array}$ & $\begin{array}
    {c} 0.45 \\ 0.5 \\ 0.05 \end{array}$ & $\begin{array}{cc}-5.02 & 1\\ -0.00 & 1.01\\ 0.51 & 1.02 \end{array}$ & $\begin{array}
    {c} 0.44 \\0.50 \\ 0.08
  \end{array}$ & $\begin{array}
    {c} \wh\beta: -0.01\\ \scriptstyle{(-0.03,0.02)}\\
    \wh\gamma: 0.10\\ \scriptstyle{(-0.07,0.12)}
  \end{array}$ & $\begin{array}
    {cl}
    \wh K: & \; 3\\ & \scriptstyle{(3,3)}\\
    \wh K_+: & \; 3\\ & \scriptstyle{(3,3)}
  \end{array}$\\
  \hline
\end{tabular}
\caption{{\footnotesize Dynamic case. Results of a Monte Carlo exercise with $100$ iterations. Study of the impact when we have two components of $\theta_{\alpha}$ very closed and/or different weights for each mixture component. $\beta^{\star} = 0$, $\gamma^{\st} = 0.1$, $K^{\star} = 3$. The estimation are means across the $100$ MC iterations of the posterior means. The credible intervals (CI) for $\beta$ and $\gamma$ are the $95\%$ CI, and the $1^{st}$ and $3^{rd}$ quartiles for $\wh K$ and $\wh K_+$.}}\label{tab:2:dynamic:MC}
\end{table}
}
\normalsize

{\footnotesize
\begin{table}[!h]
\centering
\begin{tabular}{|l|c|c||c|c|c|c||}
  \hline
  N & $\theta_{\alpha}^{\star},\theta_{\sigma^2}^{\star}$ & $w^{\star}$  & $\wh\theta$ & $\wh w_K$  & $\wh\beta$, $\wh\gamma$, & $\widehat{K} $, $\widehat{K}_+ $\\
  \hline
  \hline
  $50$ & $\begin{array}{cc}-20 & 1\\ -15 & 1\\ -10 & 1\\ -5 & 1\\ 0 & 1\\ 5 & 2 \\ 10& 0.5\\ 15 & 0.5\\ 20 & 0.5\end{array}$ & $\begin{array}
    {c} 0.11 \\ 0.11 \\ 0.11 \\ 0.11 \\ 0.11 \\ 0.11 \\ 0.11 \\ 0.11 \\ 0.11
  \end{array}$ & $\begin{array}{cc}-23.69 & 1.95\\ -17.79 & 1.97\\ -11.88 & 1.72\\ -5.76 & 2.13\\ 0.46 & 2.18\\ 6.57 & 2.46 \\ 12.76 & 1.58\\ 18.62& 1.39\\ 24.57 & 1.58
  \end{array}$ & $\begin{array}
    {c} 0.80 \\ 0.24 \\ 0.10 \\ 0.11 \\ 0.12 \\ 0.10 \\ 0.13 \\ 0.11 \\ 0.12
  \end{array}$ & $\begin{array}
    {c} \wh\beta: -0.03\\ \scriptstyle{(-0.64,0.60)}\\
    \wh\gamma: 0.63\\ \scriptstyle{(0.44,0.75)}
  \end{array}$ & $\begin{array}
    {cl}
    \wh K: & \; 4\\ & \scriptstyle{(4,5)}\\
    \wh K_+: & \; 4\\ & \scriptstyle{(4,4)}
  \end{array}$\\
  \hline
    $500$ & $\begin{array}{cc}-20 & 1\\ -15 & 1\\ -10 & 1\\ -5 & 1\\ 0 & 1\\ 5 & 2 \\ 10& 0.5\\ 15 & 0.5\\ 20 & 0.5\end{array}$ & $\begin{array}
    {c} 0.11 \\ 0.11 \\ 0.11 \\ 0.11 \\ 0.11 \\ 0.11 \\ 0.11 \\ 0.11 \\ 0.11
  \end{array}$ & $\begin{array}{cc}-19.69 & 1.23\\ -14.77 & 1.23\\ -9.98 & 1.23\\ -4.98 & 1.24\\ -0.01 & 1.26\\ 4.98 & 2.24 \\ 9.96 & 0.75\\ 14.94 & 0.75\\ 19.93 & 0.76
  \end{array}$ & $\begin{array}
    {c} 0.12 \\ 0.12 \\ 0.11 \\ 0.11 \\ 0.11 \\ 0.11 \\ 0.11 \\ 0.11 \\ 0.11
  \end{array}$ & $\begin{array}
    {c} \wh\beta: -0.05\\ \scriptstyle{(-0.21,0.11)}\\
    \wh\gamma: 0.11\\ \scriptstyle{(0.04,0.17)}
  \end{array}$ & $\begin{array}
    {cl}
    \wh K: & \; 9\\ & \scriptstyle{(9,10)}\\
    \wh K_+: & \; 9\\ & \scriptstyle{(9,10)}
  \end{array}$\\
  \hline
\end{tabular}
\caption{{\footnotesize Dynamic case. Results of a Monte Carlo exercise with $100$ iterations. Study of the impact when we increase the number of components. $\beta^{\star} = 0$, $K^{\star} = 9$, $T = 3$. The estimation are means across the $100$ Monte Carlo iterations of the posterior means. The credible interval (CI) for $\beta$ is the $95\%$ CI, and the $1^{st}$ and $3^{rd}$ quartiles for $\wh K$ and $\wh K_+$.}}\label{tab:3:dynamic:MC}
\end{table}
}
\normalsize

\section{Additional Figures}\label{Additional:figures}
{\footnotesize

\begin{figure}[h]
    \centering
    \begin{subfigure}[b]{0.45\textwidth}
        \includegraphics[width=\textwidth]{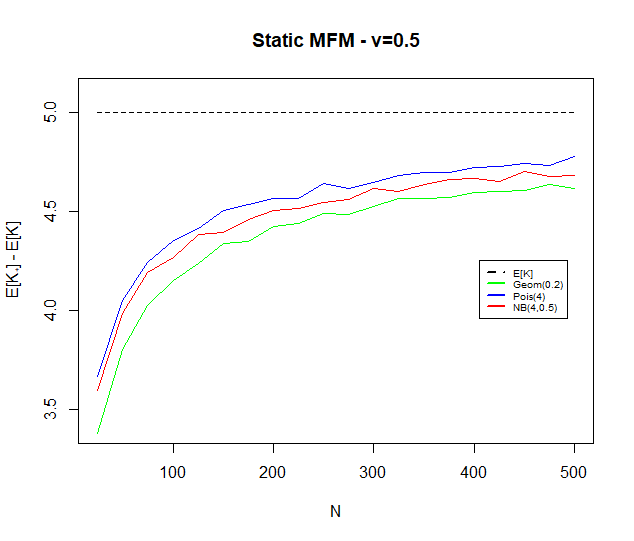}
        \caption{Static case $v=0.5$.}
        \label{fig:figure1}
    \end{subfigure}
    \hspace{0.5em}
    \begin{subfigure}[b]{0.45\textwidth}
        \includegraphics[width=\textwidth]{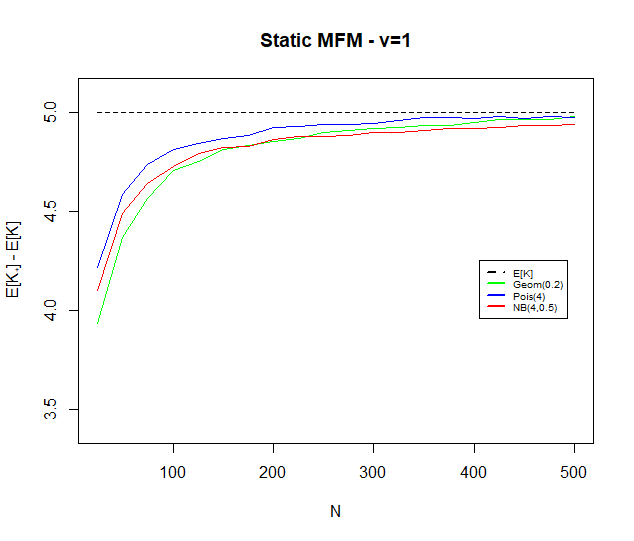}
        \caption{Static case $v=1$.}
        \label{fig:figure2}
    \end{subfigure}
    \hspace{0.5em}
    \begin{subfigure}[b]{0.45\textwidth}
        \includegraphics[width=\textwidth]{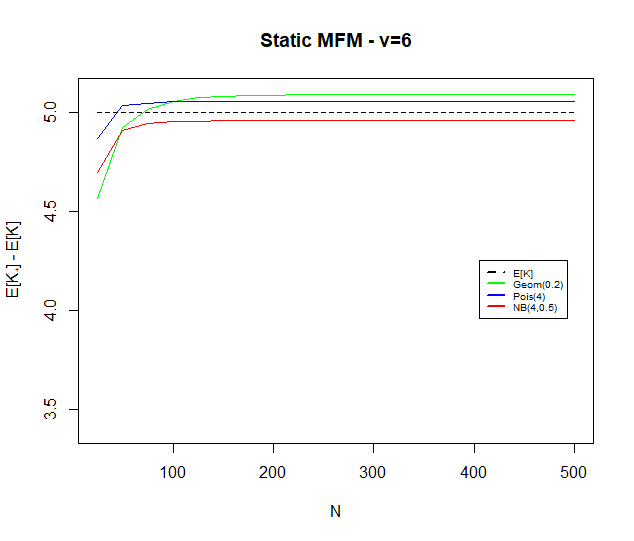}
        \caption{Static case $v=6$.}
        \label{fig:figure3}
    \end{subfigure}
%
    \begin{subfigure}[b]{0.45\textwidth}
        \includegraphics[width=\textwidth]{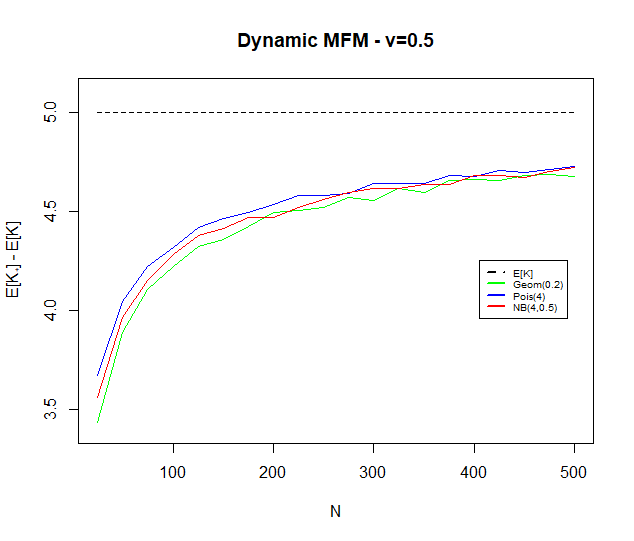}
        \caption{Dynamic case $v=0.5$.}
        \label{fig:figure1}
    \end{subfigure}
    \hspace{0.5em}
    \begin{subfigure}[b]{0.45\textwidth}
        \includegraphics[width=\textwidth]{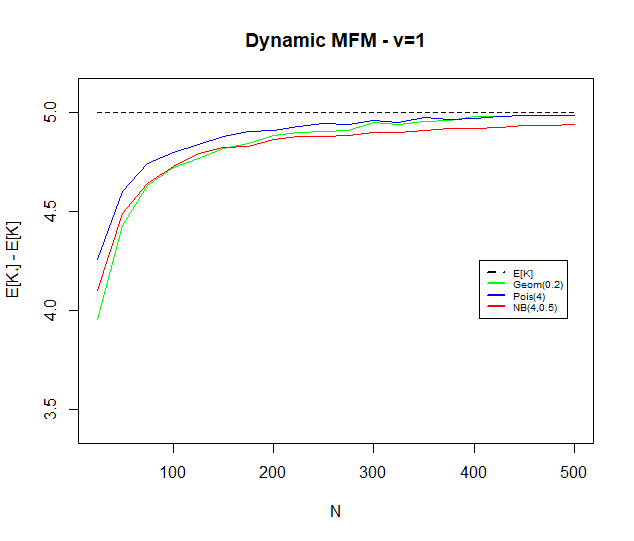}
        \caption{Dynamic case $v=1$.}
        \label{fig:figure2}
    \end{subfigure}
    \hspace{0.5em}
    \begin{subfigure}[b]{0.45\textwidth}
        \includegraphics[width=\textwidth]{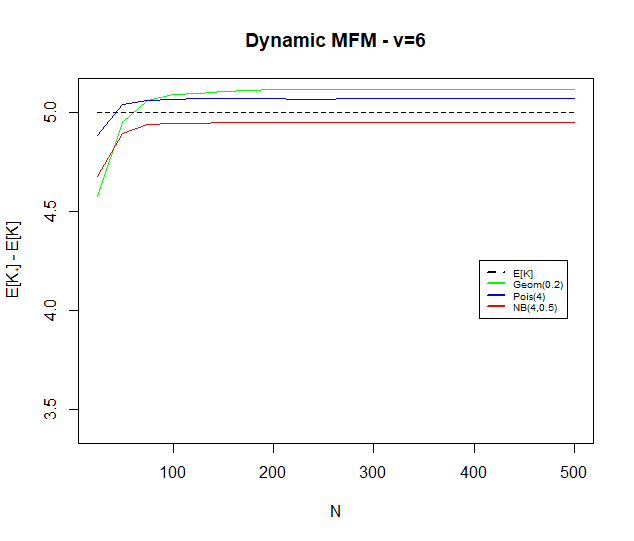}
        \caption{Dynamic case $v=6$.}
        \label{fig:figure3}
    \end{subfigure}
    \caption{{\footnotesize Effect of $N$ on the prior mean of $K_+$ for three different values of $v$. Static and Dynamic MFM and three priors for $K$: $\mathcal{G}eom(0.2)$, $\mathcal{P}oi(4)$, and $NB(4,0.5)$. We use the Geometric distribution with probability mass function $(1-p)^k p$.}}
    \label{fig:effect:N:dynamic}
\end{figure}
}
\normalsize

\begin{figure}[h]
    \centering
    \begin{subfigure}[b]{0.45\textwidth}
        \includegraphics[width=\textwidth]{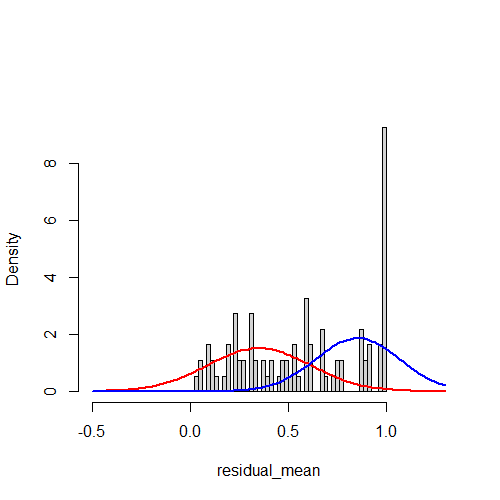}
        \caption{Histogram of the mean of $Y$ over time. $K-1\sim BNB(1,4,3)$.}
        \label{fig:figure1}
    \end{subfigure}
    \hspace{0.5em}
    \begin{subfigure}[b]{0.45\textwidth}
        \includegraphics[width=\textwidth]{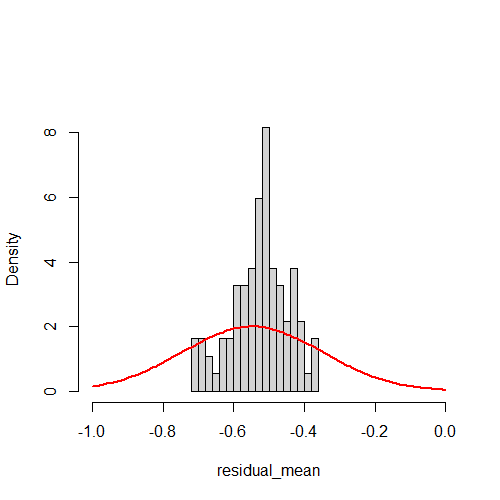}
        \caption{Histogram of residuals-mean without controls and $K-1\sim BNB(1,4,3)$.}
        \label{fig:figure2}
    \end{subfigure}
    \hspace{0.5em}
    \begin{subfigure}[b]{0.45\textwidth}
        \includegraphics[width=\textwidth]{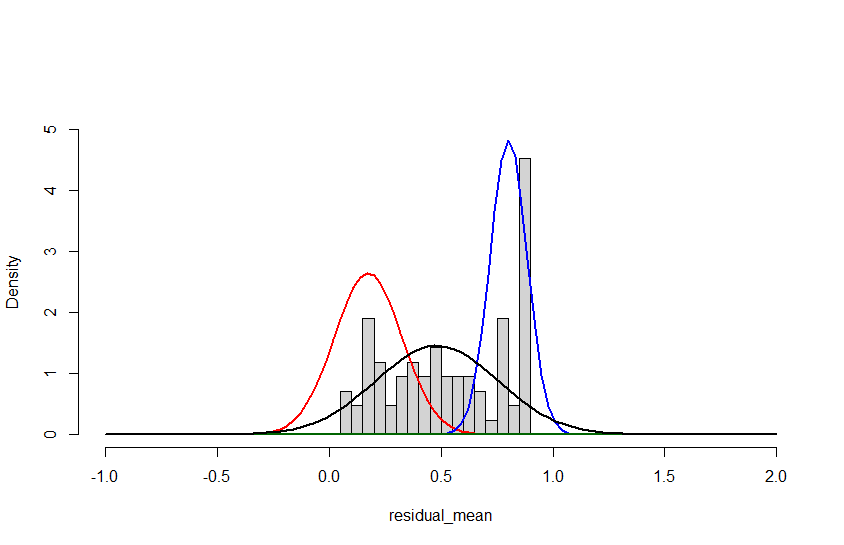}
        \caption{Histogram of residuals-mean with $age-4$ and $K-1\sim BNB(1,4,3)$.}
        \label{fig:figure3}
    \end{subfigure}
    \caption{{\footnotesize Histograms of the mean over time of the residuals from different models with and without covariates. The mean is taken over time. ``age-4'' means age group percentages (four categories) in the population plus the median age in the population.}}
    \label{fig:application:histograms}
\end{figure}

\begin{singlespace}
\bibliography{AnnaBib}
\end{singlespace}

\end{document}